\documentclass[twocolumn]{aastex631}

\usepackage{amsmath}
\usepackage{afterpage}
\usepackage{enumitem}
\usepackage{pifont}
\renewcommand\ion[2]{#1$\;${\scshape{#2}}}
\usepackage{statmath}
\usepackage{microtype}
\usepackage{graphicx}

\DeclareUnicodeCharacter{2212}{-}
\def\kms{${\text{km s}^{-1}}$}

\def\farcs{\ensuremath{\overset{\prime\prime}{.}}}

\def\arcs{\ensuremath{^{\prime\prime}}}

\def\Jyb{Jy\,beam$^{-1}$}
\def\mujy{$\mu$Jy\,beam$^{-1}$}

\def\water{H$_2$O}
\def\lsun{L$_\odot$}

\begin{document}

\title{Unveiling Ionized Jet Morphologies: Sub-arcsecond VLA Observations of Compact Radio Sources}
\shorttitle{Unveiling Ionized Jet Morphologies}

\author[0000-0003-0090-9137]{Tatiana M. Rodr\'iguez}
\affiliation{I. Physikalisches Institut der Universit\"at zu K\"oln, Z\"ulpicher Str. 77, 50937 K\"oln, Germany.}

\author[0000-0001-9942-7343]{Peter Hofner}
\affiliation{Physics Department, New Mexico Tech, 801 Leroy Pl., Socorro, NM 87801, USA.}
\affiliation{Adjunct Astronomer at the National Radio Astronomy Observatory, 1011 Lopezville Road, Socorro, NM 87801, USA.}

\author[0000-0003-3168-5922]{Emmanuel Momjian}
\affiliation{National Radio Astronomy Observatory, P.O. Box O, 1011 Lopezville Road, Socorro, NM, USA.}

\author[0000-0001-6755-9106]{Esteban D. Araya}
\affiliation{Physics Department, Western Illinois University, 1 University Circle, Macomb, IL 61455, USA.}
\affiliation{Physics Department, New Mexico Tech, 801 Leroy Pl., Socorro, NM 87801, USA.}

\author[0000-0002-9752-2125]{Ananay Sethi}
\affiliation{Physics Department, New Mexico Tech, 801 Leroy Pl., Socorro, NM 87801, USA.}

\author[0000-0001-8596-1756]{Viviana Rosero}
\affiliation{California Institute of Technology, 1200 E. California Blvd., Pasadena, CA 91125, USA.}

\begin{abstract}
We present sub-arcsecond ($\theta\sim0.1^{\prime\prime}$) resolution VLA 1.3 cm continuum and 22.2 GHz H$_2$O maser observations toward 15 compact radio continuum sources with rising spectral index and 8 string-like radio continuum structures in the \cite{Rosero16,Rosero19} survey. Three different morphologies are observed: elongated or double-peak string-like structure (6 out of 23 cases), a collection of distinct continuum peaks (4 out of 23 cases), and single compact sources (13 out of 23 cases). The majority of H$_2$O maser spots detected are within a sky-projected distance of $\sim5,000\,$au from the radio continuum peaks and tend to be well aligned and distributed in an elongated structure when more than three spots are observed. We generally recover less emission than \cite{Rosero16,Rosero19}, which together with the fact that more than half of the jet candidates in our survey appear mostly compact, suggest core/halo shock structures even on small scales. We also detected proper motion in 10 cases and measured an average projected velocity of approximately 120 km s$^{-1}$. Radio brightness variability is detected in at least two cases, possibly due to weak accretion bursts. This work, together with our previous molecular jet study, provides further evidence that support the main source of ionization in the studied sources is shocks, yet collimation is only observed in 4 cases. We conclude that the available data supports the thermal jet classification of 7 sources, and the ionized jet interpretation is further supported in 16 sources.
\end{abstract}

\keywords{\textit{Unified Astronomy Thesaurus concepts:} Jets (870); Radio jets (1347); Star-forming regions (1565); Star formation (1569); Astrophysical masers (103); Protostars (1302); Young stellar objects (1834); Pre-main-sequence stars (1290)}

\section{Introduction}
\label{sec:intro}
The details of the formation of high-mass ($>8\,M_\odot$) stars remain elusive.
The turbulent core accretion model \citep{McKee2003} suggests that high-mass stars can form in an analogous manner to low-mass stars, that is, via the gravitational collapse of a dense, turbulent core.
In this scenario, accretion and ejection processes are intrinsically connected, and bipolar jets and outflows play a key role in regulating the formation of massive young stellar objects (MYSOs) \citep[e.g.,][]{Kolligan2018,Commercon2022,Oliva2023}.
Large scale molecular outflows are ubiquitously found in high-mass star-forming regions, but ionized jets are more prominent on scales of a few 1,000\,au. At the large distances at which MYSOs are usually found, they are hardly spatially resolved. This, together with the deeply embedded nature of MYSOs, makes the study of ionized jets in high-mass star-forming regions challenging, resulting in a scarce number of these objects known. 

Relatively nearby ($<2\,$kpc) MYSOs, such as IRAS 18162$-$2048 (1.4\,kpc, \citealt{Anez-Lopez2020}) and Cep HW2 (700\,pc, \citealt{Dzib2011}), provide a great opportunity to study massive protostestellar jets with high spatial resolution. Radio observations showed that HH 80/81, powered by IRAS 18162$-$2048, is a highly collimated jet composed of several knots or shock fronts \citep{Masque2012,Masque2015}, and Cep HW2 exhibits both a wide angle wind together with a collimated jet \citep{Carrasco-Gonzalez2021}.
Even though these are only two examples, similar characteristics and morphologies have been observed in more distant MYSOs (e.g., IRAS 16562$-$3959, \citealt{Guzman2010}; IRAS 16547$-$4247, \citealt{Rodriguez2008}; G35.20$-$0.74N, \citealt{Beltran2016}; W75N, \citealt{Rodriguez-Kamenetzky2020}; IRAS 23032+5937, \citealt{Rodriguez2021}), but also in low-mass objects (e.g., HH 30, \citealt{Louvet2018}; DG Tau B, \citealt{devalon2022}; HH 212, \citealt{Lopez-Vazquez2024}; see also \citealt{Tychoniec2018}).
Furthermore, energy wise, the work by \cite{Anglada2018} shows that the empirical radio luminosity-bolometric luminosity relation derived for low-luminosity jets holds for the high-luminosity regime.
However, the high-mass end in this relation is sparsely populated, and even though recent large radio surveys such as \cite{Rosero16,Rosero19} and \cite{Purser2021} have contributed greatly in this regard, it is still needed to increase 
the sample of known jets associated with MYSOs to test and improve our current understanding of massive star formation.

\cite{Rosero16,Rosero19} carried out sensitive ($\sim3-10\,\mu$\Jyb) high angular resolution ($\theta\sim0.33$\arcs) observations at 6 and 22\,GHz toward 58 high-mass star-forming regions utilizing NRAO's\footnote{The National Radio Astronomy Observatory is a facility of the National Science Foundation operated under cooperative agreement by Associated Universities, Inc.} Karl G. Jansky Very Large Array (VLA).
The authors found that more than 30 radio continuum sources in their sample had a compact morphology and their thermal spectrum could be well modeled by a power law (i.e., an ionized jet) or by a spherical distribution of ionized gas (i.e., an \ion{H}{ii} region).
Based on additional evidence of molecular gas and/or shock tracers in the literature, they classified 13 of them as good \textit{jet candidates}. 
We carried out SiO(1$-$0) observations toward a subset of these jet candidates and found that 6 of them drive molecular jets \citep{Rodriguez23}.
The SiO abundance is highly enriched in shocked gas regions and quickly reacts when the powering source quenches, hence these detections strongly support the idea that the radio continuum emission in the 6 jet candidates studied arises from shock ionization.

In this paper we further investigate the nature of the \cite{Rosero16,Rosero19} survey sample: we present $\sim0$\farcs1 resolution VLA 22\,GHz observations of all the jet candidates. 
In Section \ref{sec:observations}, we present the sources selected for this study and provide information on the observations, data reduction, and imaging process. 
Our results are outlined in Section \ref{sec:results} and discussed in Section \ref{sec:discussion}.
A summary of this paper is provided in Section \ref{sec:summary}, and complementary figures are given in the \hyperref[sec:appendix]{Appendix}.

\vfill

\section{Observations}
\label{sec:observations}
\subsection{Sample Selection}
We selected for this work all the jet candidates from the \cite{Rosero16,Rosero19} survey, which are listed in Table 3 of \cite{Rosero19}. 
The 13 jet candidates are: UYSO1, IRAS\,18264$-$1152, IRAS\,18345$-$0641, IRAS\,18470$-$0044, IRAS\,18517+0437, IRAS\,18521+0134, G35.39$-$00.33 mm2, IRAS\,18553+0414, IRAS\,19012+0536, G53.25+00.04 mm2, IRAS\,19413+2332, IRAS\,20293+3952, and IRAS\,20343+4129.
These sources were unresolved or slightly resolved in the \cite{Rosero16} data and are potentially associated with molecular outflows reported in the literature. 
In order to enrich our sample, we also included the 8 brightest 
sources from Table 2 in \cite{Rosero19}, which have been classified by the authors as jets based on their elongated or string-like morphology, or their association with molecular outflows reported in the literature. 
These sources are G11.11$-$0.12P1, IRAS\,18151$-$1208, IRAS\,18182$-$1433, IRDC\,18223$-$3, G23.01$-$0.41, IRAS\,18440$-$0148, IRAS\,18566+0408, and IRAS\,19035+0641.
We included two additional sources in our sample that were not classified as jet candidates, but are unresolved or slightly resolved in \cite{Rosero16,Rosero19}, have a rising spectral index, and are located in infrared dark clouds. These two sources are G34.43+00.24 mm1 and G53.11+00.05 mm2. 
G34.43+00.24 mm1 was classified as hot molecular core and recent observations have shown it is located at the base of two molecular outflows \citep{Isequilla21}. 
Meanwhile, G53.11+00.05 mm2 A was the only radio continuum source in a cold molecular core detected by \cite{Rosero16,Rosero19}, which emphasizes the early stage of this MYSO. 
In total, 23 high-mass star-forming regions were observed.

\subsection{VLA K-band Observations}
We used the VLA to observe the radio continuum emission in \textit{K-}band ($18.0-26.5\,$GHz) toward the 23 target regions presented above. 
Care was taken to avoid the H$_2$O maser band when selecting the broad band spectral windows for continuum imaging.
The observations were carried out between March 7 and June 6, 2022 in the A configuration and utilized the 3-bit samplers.
For the selected frequency and array configuration, the primary beam size is $\sim2^\prime$ and the largest angular scale is 2\farcs4.
A total of 57 broad spectral windows with $128\times1\,$MHz channels were used for the continuum observations.
Based on the intensity measured by \cite{Rosero16}, the targets were classified as strong (G23.01$-$0.41, G34.43+00.24 mm2, IRAS\,19012+0536, and IRAS\,19035+0641), intermediate (IRAS~sources 18345$-$0641, 18440$-$0148, 18517+0437, 18521+0134, and 18553+0414), or weak (UYSO1, G11.11$-$0.12P1, G35.39-00.33 mm2, G53.11+00.05 mm2, G53.25+00.04 mm2, IRDC 18223$-$3, and the IRAS\,sources 18151$-$1208, 18182$-$1208, 18264$-$1152, 18470$-$0044, 18566+0408, 19413+2332, 20293+3952, and 20343+4129) if their expected intensity at 1.3\,cm in the A-configuration was $>350\,$\mujy, $<350\,$\mujy\ and $>150\,$\mujy, or $<150\,$\mujy, respectively.
The typical on-source time for strong, intermediate, and weak targets was 10, 15, and 70 minutes, respectively. 
For flux density scale and bandpass calibration, we used observations of 3C\,286. 
Complex gain calibration was performed using nearby calibrators to each target source, as shown in Table \ref{tab:calibrators}, with a cycle time of 4 minutes.

We utilized NRAO's Common Astronomy Software Applications (CASA, \citealt{casa}) Calibration Pipeline version 6.2.1.7 to process the radio continuum data. 
The images were made using the CASA task \texttt{tclean} with a robust parameter of 0.5, resulting in a typical synthesized beam size of approximately 0\farcs08. 
The observed regions and their phase centers, as well as the synthesized beam sizes and rms noise values of the final continuum images are listed in columns 1 through 6 of Table \ref{tab:regions-images}. In columns 7 and 8, we also list the distance and luminosity of the regions.
The central frequency of the final continuum images is 22.2\,GHz, and we will refer to it as 1.3\,cm continuum hereafter.

We simultaneously observed the 22.23508\,GHz H$_2$O ($6_{1,6}-5_{2,3}$) maser transition. 
A narrow (16\,MHz) spectral window with $1024 \times 15.6\,$kHz channels was used for the H$_2$O maser observations, which resulted in a channel spacing of about 0.2\,\kms. 
Details on the calibration and imaging of the H$_2$O masers will be provided in a follow up work.


\begin{deluxetable}{ll}
\tablenum{1}
\tabletypesize{\scriptsize}
\tablecaption{Observed Complex Gain Calibrators}
\label{tab:calibrators}
\tablecolumns{2}
\tablewidth{0pt}
\tablehead{
\colhead{Calibrator} & \colhead{Target source(s)}
}
\startdata
J0730$-$1141 & UYSO1. \\
J1820$-$2528 & G11.11$-$0.12\,P1. \\
J1832$-$1035 & 18151$-$1208, 18182$-$1433, IRDC 18223$-$3, 18264$-$1152,\\
    & G23.01$-$0.41, 18345$-$0641. \\
J1851+0035 & 18440$-$0148, 18470$-$0044, G34.43+00.24 mm1, \\
    & 18517+0437, 18521+0134, G35.39$-$00.33 mm2, \\
    & 18553+0414, 18566+0408, 19012+0536, 19035+0641. \\
J1925+2106 & G53.11+00.05 mm2, G53.25+00.04 mm2, 19413+2332. \\
J2015+3710 & 20293+3952, 20343+4129.\\
\enddata
\end{deluxetable}

\begin{deluxetable*}{l c c L R C C C}[htpb]
\tablenum{2}
\label{tab:regions-images}
\tabletypesize{\scriptsize}
\tablecaption{Observed regions.}
\tablecolumns{8}
\tablewidth{0pt}
\tablehead{
\colhead{Region} & \colhead{RA$_{\text{J2000}}$} & \colhead{Dec$_{\text{J2000}}$} & \multicolumn{2}{c}{$\theta_{syn}$} & \colhead{$\sigma$} & \colhead{$d$} & \colhead{$L_\mathrm{bol}$} \\
\colhead{} & \colhead{(h m s)} & \colhead{($^\circ \, ^\prime \, ^{\prime\prime}$)} & \colhead{($^{\prime\prime}\, \times\, ^{\prime\prime}$)} & \colhead{$(^\circ)$} & \colhead{(\mujy)} &  \colhead{(kpc)}  & \colhead{(log \lsun)}
}
\startdata
UYSO1 &  07 05 10.80 & $-$12 18 59.0 & 0.12\times0.08 &  12.42  &  4.78 & 1.0 & 0.2   \\
G11.11$-$0.12P1 &  18 10 28.40 & $-$19 22 30.0 & 0.14\times0.07 & $-$20.92 & 4.97 & 2.9 & 3.1 \\
IRAS 18151$-$1208 &  18 17 58.30 &  $-$12 07 24.0 & 0.13\times0.08  &  $-25.06$  &  5.01 & 2.0 & 3.9 \\
IRAS 18182$-$1433 & 18 21 07.90 & $-$14 31 53.0 & $0.13 \times 0.07$ & 16.77 & 5.99 & 3.6& 4.1  \\
IRDC 18223$-3$  & 18 25 08.50 &  $-$12 45 23.0 &  $0.13 \times 0.08$ &  22.25 & 5.92 & 3.7   & 0.6  \\ 
IRAS 18264$-$1152  &  18 29 14.42 & $-$11 50 24.3 & $0.12 \times 0.08$& $-$19.10 & 4.59 &3.3 & 3.9\\
G23.01$−$0.41 &   18 34 40.28 & $-$09 00 38.0 & $0.11 \times 0.08$& $-$13.62 & 10.68 & 4.6& 4.5 \\
IRAS 18345$-$0641 &  18 37 16.90 & $-$06 38 32.0 & $0.11 \times 0.08$& $-$13.12 & 4.50 & 9.5 & 5.3\\
IRAS 18440$-$0148 &  18 46 36.61 & $-$01 45 23.0 & $0.10 \times 0.08$ & -28.92 & 9.52 & 5.2 & 3.5 \\
IRAS 18470$-$0044 &  18 49 37.75 & $-$00 41 00.2 & 0.09 \times 0.08 & 0.95 & 4.45 & 6.5 & 4.5\\
G34.43+00.24 mm1 & 18 53 18.01 & +01 25 24.0 & $0.09 \times 0.08$ & $-$34.96 & 11.94 & 9.8 & 5.1\\
IRAS 18517+0437  & 18 54 13.80 &	+04 41 32.0 & $0.09\times0.08$ & 9.91 & 8.93 & 1.9 & 3.8 \\
IRAS 18521+0134 &18 54 40.80 & +01 38 02.0 & $0.09\times0.08$ & 10.93 & 8.55 &9.1 & 4.5 \\
G35.39$-$00.33 mm2 & 18 56 59.06& +02 04 54.6 & $0.09\times0.07$ & $-$2.98 & 5.30 & 2.3 & 2.8  \\
IRAS 18553+0414 & 18 57 52.90 &	+04 18 16.0 & $0.10\times 0.08$ & $-$29.25 & 7.82 & 12.3 & 4.8\\
IRAS 18566+0408 & 18 59 10.03 & +04 15 15.5 & $0.10\times0.08$ & $-$33.20 & 4.01 & 6.7& 4.7 \\
IRAS 19012+0536 & 19 03 45.26 &	+05 40 40.0 & $0.10\times0.08$ & $-$41.37 & 9.53 & 4.2& 4.0  \\
IRAS 19035+0641 & 19 06 01.10 & +06 46 35.0 & $0.08\times0.07$ & 1.56 & 13.73 & 10.8 & 5.1\\
G53.11+00.05 mm2 & 19 29 20.20 & +17 57 06.0 & $0.08\times0.07$ & 5.64 & 4.90 & 1.9 & 1.9 \\
G53.25+00.04 mm2 & 19 29 34.50 & +18 01 39.0 & $0.08\times0.07$  & 4.70 & 7.24 & 2.0  & 2.1 \\
IRAS 19413+2332 & 19 43 28.90 & +23 40 04.0 & $0.29\times0.25^\dagger$ & 58.64 & 7.5 & 1.8/6.8 & 2.9/4.1 \\
IRAS 20293+3952  & 20 31 13.35 &	+40 03 11.3 & $0.10\times0.08$ & $-82.97$ & 6.40 & 1.3/2 & 3.1/3.5  \\
IRAS 20343+4129  & 20 36 07.52 & +41 40 09.1 & $0.10\times0.08$  & $-84.00$ & 6.64 &  1.4 & 3.0 \\  
\enddata
\tablenotetext{}{$^\dagger$ Tapering was applied to image the extended emission.}
\end{deluxetable*}

\section{Results}
\label{sec:results}

\subsection{Radio continuum}
We detected 1.3\,cm continuum emission toward all the target regions except for IRDC\,18223$-$3. The radio continuum images can be found in the \hyperref[sec:appendix]{Appendix}.
In columns 1 and 2 of Table \ref{tab:cont-sources}, we list the radio continuum peaks detected following the nomenclature used by \cite{Rosero16}.
Radio continuum features reported by \cite{Rosero16} that we did not detect are marked with an asterisk. 
Those continuum peaks that we detect but are not reported or detected by \cite{Rosero16} are marked with a dagger and named from East to West using a letter in alphabetical order and following the last letter used by the authors in the region. 
The radio continuum sources from \cite{Rosero16} that are found to be composed of two or more continuum peaks in our data are marked with a double dagger symbol and a natural number was added to the letter denomination, starting from 1 and increasing from East to West. 
In columns 3 and 4 of Table \ref{tab:cont-sources}, we list each radio continuum peak position.
In column 5 we indicate whether the radio continuum source presents an elongated or string-like structure (S) and if it is associated with shock tracers from this or our previous work, that is, if it appears spatially connected to at least two H$_2$O maser spots (W) or if it has been associated with a molecular jet in \cite{Rodriguez23} (MJ). 
We also note if it can be associated with a molecular outflow or other jet/shock tracer data reported in the literature (L) and provide the relevant references in the last column of the table.

With a slightly better rms noise than \cite{Rosero16}, we were not able to detect emission $>3\sigma$ toward IRDC\,18223$-$3.
This was one of the weaker sources in our sample, with a peak intensity at 1.3\,cm of about 40\,\mujy\ in \cite{Rosero16}, and this non-detection might be due to a combination of sensitivity, our smaller synthesized beam size, and jet brightness variability.
Similarly, IRAS\,19413+2332 A was not detected in our high angular resolution observations, but a $\sim4.5\sigma$ source was recovered after a slight taper was applied (see Table \ref{tab:regions-images} and the top panel of Figure \ref{fig:appendix-A10}). 
Lastly, we also report on G53.25+00.04 mm4, which with a compact morphology and a moderately rising spectral index ($\alpha=0.1$) bared no classification by \cite{Rosero16,Rosero19}.

\startlongtable
\begin{deluxetable*}{l l c c  c C}
\tablenum{3}
\label{tab:cont-sources}
\tabletypesize{\scriptsize}
\tablecaption{Radio sources in our sample.}
\tablecolumns{6}
\tablewidth{0pt}
\tablehead{
\multicolumn{2}{c}{Source} & \colhead{RA$_{\text{J2000}}$} & \colhead{Dec$_{\text{J2000}}$} & \colhead{Criteria$^a$} & \colhead{Ref.}\\
\colhead{} & \colhead{} & \colhead{(h m s)} & \colhead{($^\circ \, ^\prime \, ^{\prime\prime}$)}  & \colhead{} & \colhead{}
}
\startdata
UYSO1 & A & 07 05 10.939 & $-$12 19 00.47 &  L    & (1) \\
& B &  07 05 10.810 & $-$12 19 56.76 &        & \\
\hline
G11.11$-$0.12P1 & A & 18 10 28.395 & $-$19 22 29.91 & S, L & (2), (3)\\
& B & 18 10 28.329 & $-$19 22 30.55 &      & \\
& C$^\ast$ &  $-$  &   $-$    &     & \\
& D$^\ast$ & $-$ &  $-$   &    & \\
& E$^\dagger$ &  18 10 27.920  & $-$19 22 36.38 &      & \\
\hline
IRAS 18151$-$1208 & A   &  18 17 58.333  &  $-$12 07 23.99&    & \\
  & B &  18 17 58.123  &  $-$12 07 24.77 &L & (4)$-$(9)\\
 & C$^\ast$ & $-$  & $-$  &   &    \\
 & D$^\dagger$&  18 17 58.046 &  $-$12 07 23.06  &   &     \\
\hline
IRAS 18182$-$1433 & A$^\ast$ & $-$ & $-$ &  &  \\
& B$^\ast$ & $-$ & $-$ & &   \\
& C1$^\ddagger$ & 18 21 09.124  & $-$14 31 48.68 &  W, L  & (9)$-$(13)  \\
& C2 &  18 21 09.109  & $-$14 31 48.03 &  &    \\
\hline
IRDC 18223$-3$  &  A$^\ast$ & $-$&  $-$   &   L & (14), (15) \\ 
  &  B$^\ast$ & $-$&  $-$   &  &    \\
\hline
IRAS 18264$-$1152  &  A & 18 29 14.685 & $-$11 50 23.60  & &\\
  &  B$^\ast$ & $-$ & $-$ &    &  \\
  &  C$^\ast$ & $-$ & $-$ &   &   \\
  &  D & 18 29 14.431 & $-$11 50 23.70 &   &    \\
  &  E &  18 29 14.417 & $-$11 50 24.47 &   &    \\
 & F1$^\ddagger$ & 18 29 14.360 & $-$11 50 22.62 &  W, L   &  (4), (6), (7), (9), (12), (13), (16), (17) \\
  &  F2 & 18 29 14.355 & $-$11 50 22.40 &  &    \\
  &  F3 & 18 29 14.338 & $-$11 50 22.50 &   &  \\
  &  F4 & 18 29 14.306 & $-$11 50 22.48  &   &   \\
&  G &  18 29 14.223 & $-$11 50 22.61 &  &  \\
&  H &  18 29 14.342 & $-$11 50 20.09 &    &  \\
&  I$^\dagger$ & 18 29 14.288 & $-$11 50 20.64 &   &   \\
\hline
G23.01$−$0.41 &  A  & 18 34 40.283	& $-$09 00 38.33  & W, L & (18)$-$(20) \\
\hline
IRAS 18345$-$0641 & A1$^\ddagger$ & 18 37 16.921 & $-$06 38 30.60 &  MJ, L & (4), (7), (9), (21) \\
& A2 &  18 37 16.900  &  $-$06 38 30.29 &   &   \\
\hline
IRAS 18440$-$0148 & A & 18 46 36.603 & $-$01 45 22.70  & L & (22) \\
    & B$^\dagger$ & 18 46 36.417 & $-$01 45 22.15  &  &  \\
\hline
IRAS 18470$-$0044 & B & 18 49 37.747 & $-$00 41 00.25 &  L & (4) \\
 & C & 18 49 37.760 & $-$00 41 01.41 &    &   \\
\hline
G34.43+00.24 mm1 & A1$^\ddagger$ & 18 53 18.014 & +01 25 25.53  &  S, W, L  & (23)\\
& A2 &  18 53 18.006  &  +01 25 25.47 &   &   \\
\hline
IRAS 18517+0437 &  A  & 18 54 14.243 &	+04 41 40.84 & W, MJ, L & (6), (7), (9), (12), (21) \\
\hline
IRAS 18521+0134 &  A  &18 54 40.739 & +01 38 06.46  &  W, L & (9)\\
\hline
G35.39$-$00.33 mm2  &  A & 18 56 59.058 &	+02 04 54.47  &  W & \\
\hline
IRAS 18553+0414 &  A1$^\ddagger$   & 18 57 53.380 &	+04 18 17.35 &   S, W, MJ, L  & (9), (21) \\
& A2 & 18 57 53.368   &   +04 18 17.28  &    &   \\
\hline
IRAS 18566+0408 & A1$^\ddagger$ & 18 59 10.037 & +04 12 15.44 & S, W, L & (4), (9), (24)$-$(27)\\
& A2 &18 59 10.015 & +04 12 15.58 &     &   \\
& B &  18 59 09.966  &  +04 12 15.55  &   &  \\
& C$^\ast$ &  $-$  &   $-$    &   & \\
& D$^\ast$ &  $-$  &   $-$    &   &   \\
\hline
IRAS 19012+0536 &  A  & 19 03 45.265 &	+05 40 42.68 &   W, MJ, L & (4), (21) \\
\hline
IRAS 19035+0641 & A1$^{\S}$ & 19 06 01.625 & +06 46 36.74  & S, W, L  & (4), (6), (29) \\
 & A2$^{\ddagger}$  & 19 06 01.605&	+06 46 36.17 &   & \\
& A3 & 19 06 01.599 & +06 46 36.18 &    &   \\
\hline
G53.11+00.05 mm2 & A1$^\ddagger$  & 19 29 20.666 & +17 57 18.16 &  MJ & (21) \\
& A2 & 19 29 20.649 & +17 57 18.17 &    &     \\
\hline
G53.25+00.04 mm2 &  A & 19 29 33.525 & +18 00 54.15 &   N  & \\
\hline
G53.25+00.04 mm4 &  A & 19 29 34.200 & +18 01 39.08 &   W  & \\
\hline
IRAS 19413+2332 & A & 19 43 29.829 & +23 40 23.45  & N$^b$ & (4) \\
\hline
IRAS 20293+3952 &  E  & 20 31 12.890 &	+40 03 22.69 &  MJ, L & (7), (9), (21), (30) \\
\hline
IRAS 20343+4129  &  B   & 20 36 07.521 & +41 40 09.02 & L & (31), (32)\\
\enddata
\tablenotetext{}{$^a$ Indicates whether the radio continuum emission presents an elongated or string-like structure (S), appears spatially connected to at least two \water\ maser spots (W), is associated with a molecular jet as per \cite{Rodriguez23} (MJ), can be associated with a molecular outflow or other jet shock tracer data reported in the literature (L), or if none of these are fulfilled (N).\\ $^b$ CO emission detected in the region but no clear bipolar structure is seen.}
\tablenotetext{}{$^\ast$ The source was not detected in our observations.}
\tablenotetext{}{$^\dagger$ The source was not reported in \cite{Rosero16}.}
\tablenotetext{}{$^\ddagger$ The 1.3 cm continuum source from \cite{Rosero16} was resolved into more than one feature in our observations.}
\tablenotetext{}{$^\S$ We adopted the radio feature nomenclature from \cite{Rodriguez2024}.}
\tablenotetext{}{References: (1) \cite{Forbrich09}, (2) \cite{Wang14}, (3) \cite{Rosero2014}, (4) \cite{Beuther02}, (5) \cite{Davis04}, (6) \cite{LopezSepulcre10}, (7) \cite{Varricatt10}, (8) \cite{Fallscheer11}, (9) \cite{RG17}, (10) \cite{Beuther06}, (11) \cite{Sanna10b}, (12) \cite{Liu22}, (13) \cite{Taniguchi23}, (14) \cite{Beuther05}, (15) \cite{Fallscheer09}, (16) \cite{CostaSilva22}, (17) \cite{Saha22}, (18) \cite{Sanna10}, (19) \cite{Sanna14}, (20) \cite{Sanna16}, (21) \cite{Rodriguez23}, (22) \cite{Sridharan02}, (23) \cite{Isequilla21}, (24) \cite{Araya07}, (25) \cite{Zhang07} , (26) \cite{Hofner17}, (27) \cite{Silva17}, (28) \cite{Zhang17}, (29) \cite{Rodriguez2024}, (30) \cite{Beuther04}, (31) \cite{Kumar02}, (32) \cite{Palau07}.}
\end{deluxetable*}

\begin{figure*}
    \centering
    \includegraphics[width=.45\textwidth]{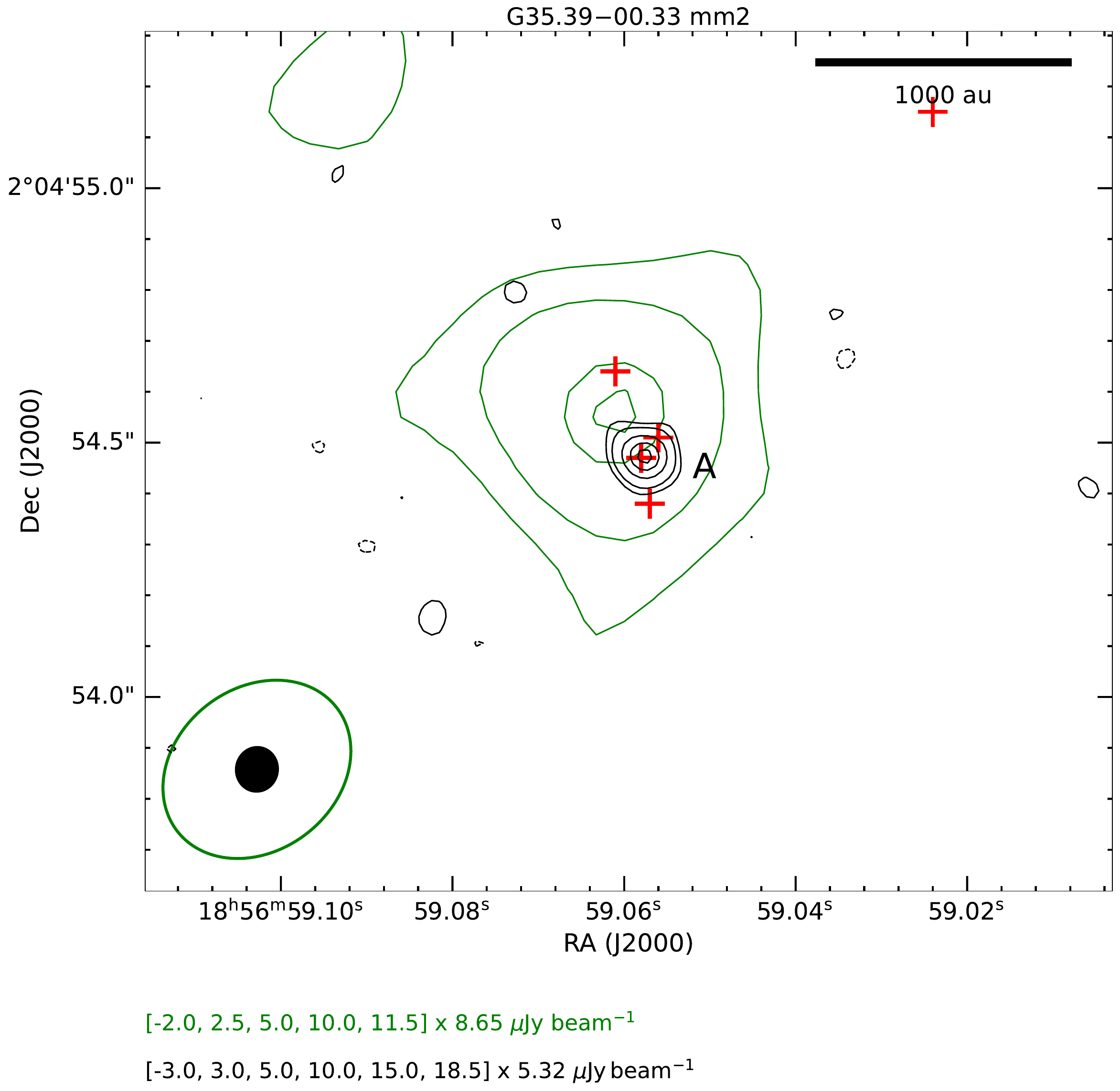}
    \includegraphics[width=.45\textwidth]{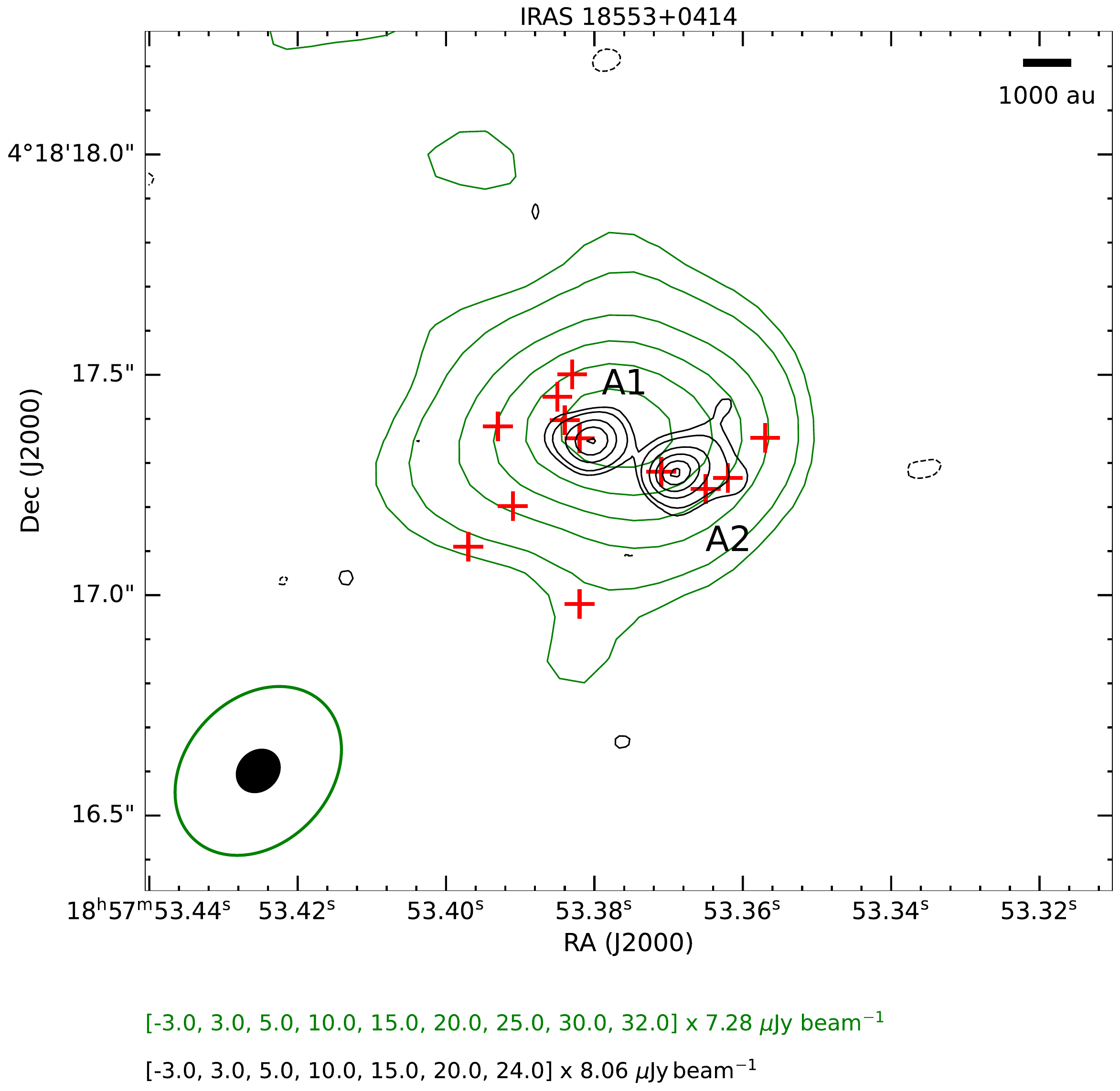}
    \includegraphics[width=.45\textwidth]{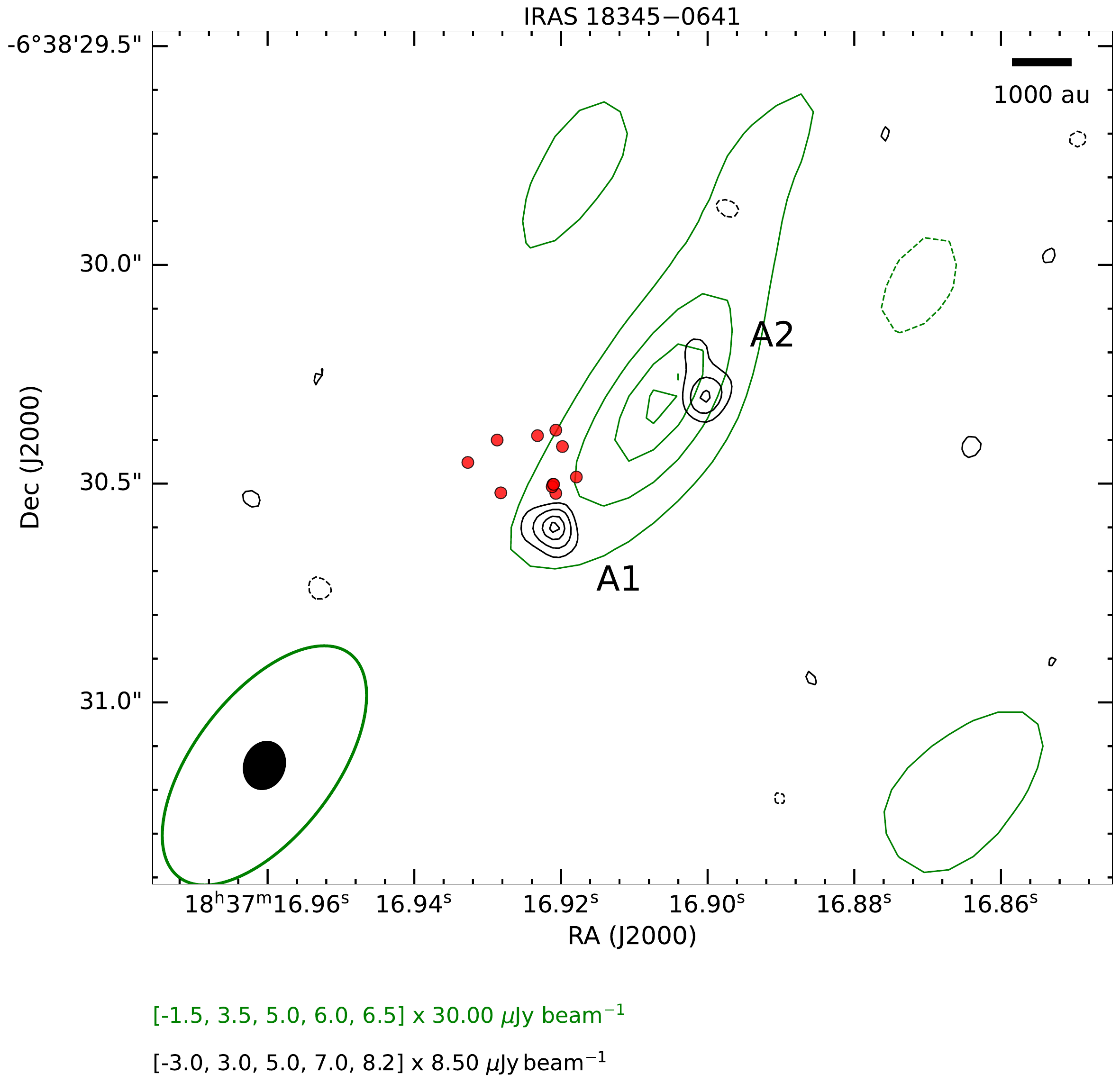}
    \includegraphics[width=.45\textwidth]{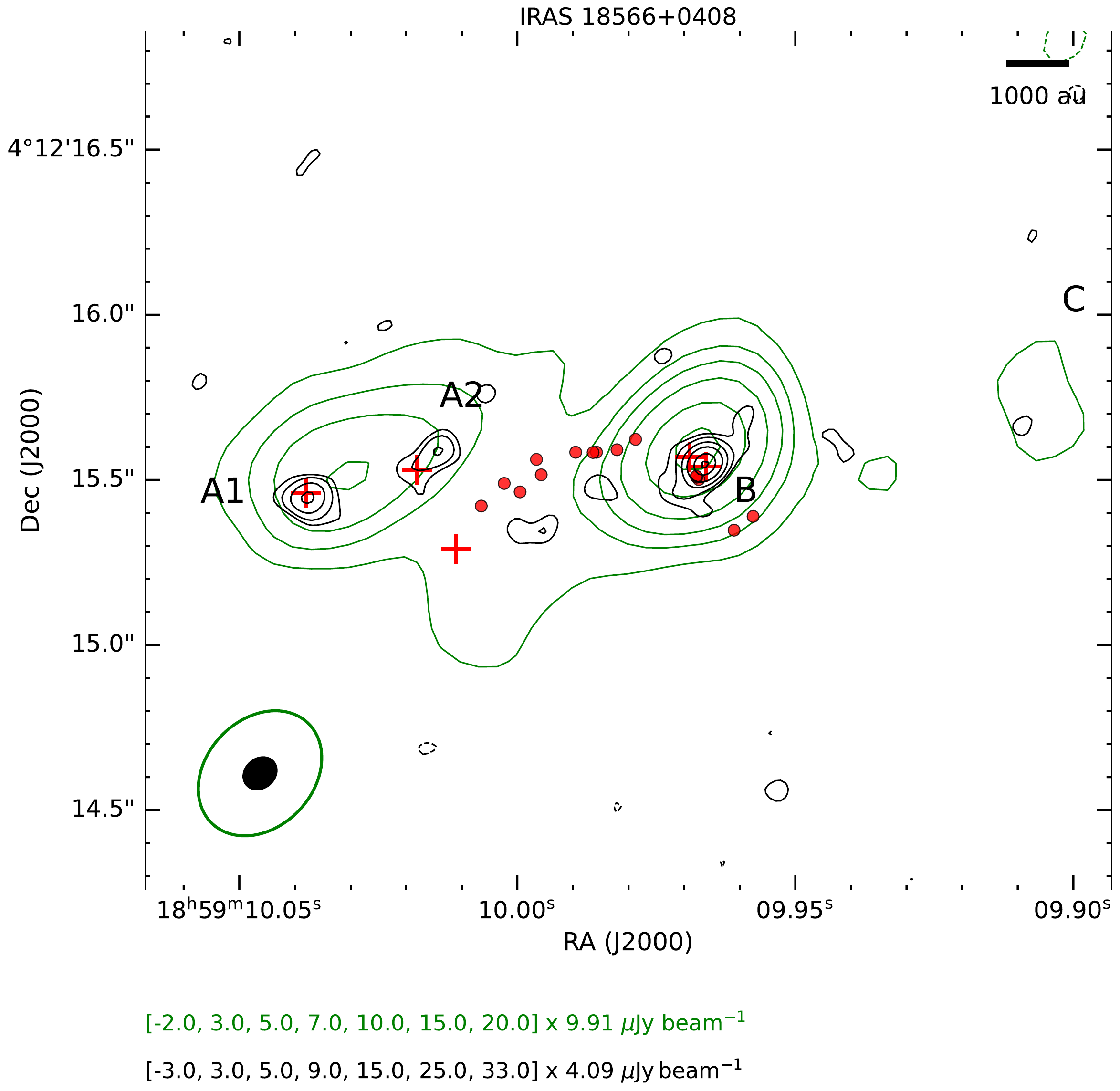}
    \caption{Examples of the different morphologies observed:
    compact (top left, G35.39$-$00.33 mm2), double-peaked with lower level continuum connection (top right, IRAS\,18553+0414), two or more individual compact features (bottom left, IRAS\,18345$-$0641), and string-like (bottom right, IRAS\,18566+0408).
    In every panel, the black and green contours represent the 1.3\,cm continuum emission from our observations and those of \cite{Rosero16}, respectively. The red $+$ symbols mark the position of the \water\ masers from this work, and the red circles in IRAS\,18345$-$0641 and IRAS\,18566+0408 show the position of the 6.7 GHz CH$_3$OH masers reported by \cite{Bartkiewicz09} and \cite{Araya10}, respectively. The synthesized beam sizes are represented by the ellipses in the bottom left corner of each panel.
    }
    \label{fig:examples}
\end{figure*}

Regarding the observed morphologies, we find that the radio continuum sources that
appeared as a single, compact 1.3 cm continuum structure (i.e., no clear observed substructure, and the shape and size are comparable to those of the synthesized beam) in \cite{Rosero16,Rosero19}:
\begin{itemize}
    \item remain mostly compact in 13 out of 19 cases and are located within a synthesized-beam-size ($\sim$0\farcs33) from the radio continuum peak from \cite{Rosero16}. These are: UYSO1 A, IRAS\,18521+0134 A, G23.01$-$0.41 A, IRAS\,18440$-$0148 A, IRAS 18470$-$0044 B, IRAS 18151$-$1208 B, IRAS\,18517+0437 A, G35.39$-$00.33 mm2 A, IRAS\,19012+0536 A, G53.25+00.04 mm2 A, G53.25+00.04 mm4 A, IRAS\,20293+3952 E, and IRAS\,20343+4129 B (see top left of Figure \ref{fig:examples} and the respective \hyperref[sec:appendix]{Appendix} images).
    Note that this classification includes both unresolved sources and single resolved sources that show a slight elongation in at least one direction.    
    
    \item present an elongated and/or double-peaked structure, i.e., two or more distinct peaks with a lower level ($\gtrsim3\sigma$) connection between them, in 4 out of 19 cases: IRAS\,18264$-$1152, G34.43+00.24 mm1, IRAS\,18553+0414, and IRAS\,19035+0641 (see top right panel of Figure \ref{fig:examples} and the respective \hyperref[sec:appendix]{Appendix} images). 
    
    \item consist of two individual, well separated, compact structures in 2 out of 19 cases: IRAS\,18345$-$0641 and G53.11+00.05 mm2 (see bottom left panel of Figure \ref{fig:examples} and the respective \hyperref[sec:appendix]{Appendix} images).
    In these cases, the linear separation between the peaks are 570 (G53.11+00.05 mm2 A1 and A2) and 3,800\,au (IRAS\,18345$-$0641 A1 and A2). 
\end{itemize}
We remark that most of the sources in our sample appear as compact structures at 0\farcs08 resolution, be it a single feature (for example, G35.39$-$00.33 mm2) or multiple (for example, IRAS 18345$-$0641).
Furthermore, we recover $\sim50-97$\% of the emission reported in \cite{Rosero16}, which shows that the bulk of the emission remains compact.

The 3 radio continuum sources studied here that exhibit a string-like morphology in the \cite{Rosero16,Rosero19} survey (G11.11$-$0.12P1, IRAS\,18182$-$1433 C, and IRAS\,18566+0408) appear as a collection of two or more mostly compact sources in our higher angular resolution observations.
These compact structures are found toward the peak of the \cite{Rosero16} 1.3\,cm emission and appear well aligned (e.g., bottom right panel of Figure \ref{fig:examples}).
We note, however, that we do not detect all the features of these elongated or string-like sources. 
The most notable case is IRAS\,18182$-$1433, where we observe source C (named C1 and C2 in this work), but the 1.3 cm peaks B, D and E from \cite{Rosero16,Rosero19} are not detected.
Moreover, IRAS\,18182$-$1433 C in \cite{Rosero16,Rosero19} exhibits an elongation not only in the NW direction (which we resolve into source C2), but also in the EW direction. 
We note that this component, which is not detected in our observations, is non-thermal and it most likely traces the jet driven by source C1 \citep[see][]{Moscadelli2013}.

We obtained 2-D Gaussian fits toward all the individual 1.3\,cm continuum peaks detected.
In Table \ref{tab:cont-fit}, we list the obtained deconvolved sizes, in units of milliarcseconds (mas) and au, and the position angle (P.A.), as well as the measured peak intensity and flux density. 
The emission was successfully deconvolved from the synthesized beam in 20 out of 38 cases.
The smallest and largest deconvolved sizes obtained were toward the closest and furthest sources respectively, and are $75\times65\,$au$^2$ (UYSO1\,A) and $955\times640\,$au$^2$ (IRAS\,18553+0414\,A2). The average deconvolved size is approximately $280\times 120\,$au$^2$.
Remarkably, we are able to trace linear structures in the range of a few tens to a few hundred au, comparable to the size of the Solar System.
We were able to deconvolve the emission in only one direction for 9 radio continuum features, and we were unable to deconvolve the remaining 10 peaks. 
The average upper limit size obtained in these cases is approximately $300\times 175\,$au$^2$.

\subsection{\texorpdfstring{$H_2O$}{H2O} masers}
We detected \water\ masers in 19 of the observed regions, that is, in all of them except G11.11$-$0.12P1, IRAS 18345$-$0641, IRDC 18223$-$3, and IRAS 19413+2332. 
The peak position of the maser spots are marked with red crosses in Figure \ref{fig:examples} and in the \hyperref[sec:appendix]{Appendix}.
We detected between 1 and 20 distinct maser spots per region, 
which are usually observed within 5,000\,au from the brightest continuum source in the region and tend to be distributed in roughly linear structures in the cases where more than 3 spots are detected: IRAS 18182$−$1433, G23.01$−$00.41, G35.39$−$00.33 mm2, IRAS 18566+0408, IRAS 19012+0536, and IRAS 19035+0641.
Water masers arise in outflow cavities and accretion disks \citep{Moscadelli2020,Moscadelli2024}, and they are prominent until the emergence of an HII region \citep[][and references therein]{Urquhart2024}.
Accretion disks around MYSOs typically have radii of a few 1,000\,au \citep[e.g.,][]{Johnston2015,Ilee2016,Sanna2019}, thus a kinematic study including radial velocities and proper motion measurements is needed to unambiguously determine from where the detected masers arise.
A detailed study of the observed H$_2$O masers will be presented in a subsequent publication.

\begin{deluxetable*}{l l R R R R C C R}
\tablenum{4}
\label{tab:cont-fit}
\tabletypesize{\scriptsize}
\tablecaption{Results of the 2-D Gaussian fits to the detected 1.3 cm radio sources.}
\tablecolumns{9}
\tablewidth{0pt}
\tablehead{
\multicolumn{2}{c}{Source}& \multicolumn{2}{c}{$\theta_{maj}$} & \multicolumn{2}{c}{$\theta_{min}$} & \colhead{P.A.} & \colhead{$I_\nu$} & \colhead{$S_\nu$} \\
\colhead{} & \colhead{} & \colhead{(mas)} & \colhead{(au)} & \colhead{(mas)} & \colhead{(au)} & \colhead{($^\circ$)} & \colhead{($\mu\text{Jy beam}^{-1}$)} &  \colhead{($\mu\text{Jy}$)}
}
\startdata
UYSO1 & A & 75 & 75 & 65 & 65   & 50.5  &   173.0\,(4.7) &   266.5\,(11.0)\\
    & B & 88  & 88  & 74  & 74  & 112.2  & 62.6\,(4.6)  & 109.6\,(11.9) \\
G11.11$-$0.12P1 & A     &  73 &  263 &  < 72 & < 259  &   117.3 &  75.3\,(4.9)    &  93.1\,(9.8) \\
              & B &        153   & 551 &  < 40 & < 144   &  141.8  & 37.4\,(4.7)     & 64.7\,(12.0) \\
              &  E &        <95  & <341 &  < 47 & < 168   &  -  & 33.8\,(4.9)     & 33.8\,(4.9) \\
IRAS 18151$-$1208  & A &  < 66 & < 185  & < 29  & <82  & -  & 52.6\,(5.0)  &  52.6\,(5.0)  \\
                & B & 33  & 93  & 15  & 41  & 62.8  & 626.5\,(5.0)  & 682.6\,(9.2)  \\
    & D & < 73 & <205  & < 29  & <80  & -  & 43.7\,(5.0)  &  43.7\,(5.0) \\
IRAS 18182$-$1433 & C1 & 67  & 241  &  23 & 82  &  68.0   & 167.2\,(5.9)    & 215.4\,(12.2) \\
            & C2 & <63  & <228  & <34  & <123  & -  & 45.1\,(6.0)  &  45.1\,(6.0) \\
IRAS 18264$-$1152  & F1 & 74    &  243 &  16 &  54 &  123.1  & 230.9\,(4.5)    & 291.9\,(9.2) \\
            &   F2  & 97 & 321   & 22 & 74   &   173.7 & 39.4\,(4.5) & 53.0\,(9.5) \\
            &   F3  &  85 & 280  & 20 &  67  &   61.3 &  49.6\,(4.4) &   73.0\,(10.1)\\
            &   F4  &  < 79 &  < 261 & < 84 & < 277 & - & 46.4\,(4.6) & 46.4\,(4.6) \\
G23.01$−$0.41 & A & 77  & 354 &   16 &  72  &  71.9   & 368.4\,(10.4)   & 510.7\,(22.6) \\
IRAS 18345$-$0641 & A1 & <54  & <510  & < 63 & < 326  & -   & 72.1\,(8.4)     &  72.1\,(8.4) \\
             & A2 & <61 & <580 & < 31 & < 292 &  -   & 60.3\,(8.3)    & 60.3\,(8.3) \\
IRAS 18440$-$0148 & A & 73 & 380 & 11 & 59 & 138.2 & 139.4\,(9.3) & 175.5\,(18.9) \\
IRAS 18470$-$0044 & B & 36 & 234 & 23  & 149 & 77.0 & 117.6 (3.7) & 132.8 (7.1) \\
G34.43+00.24 mm1 & A1 &  74    & 725.2  & 38  & 372 &  99.4   & 493.8\,(11.5)  &  702.4\,(25.3) \\
         & A2   &  115   & 1127  & 10 & 98 &  52.4   & 125.4\,(11.3) & 214.4\,(28.5) \\
IRAS 18517+0437 &  A &  78 & 148  &    42  & 80   &  16.5   & 222.0 \,(8.7)    & 334.7\,(20.1) \\
IRAS 18521+0134 & A & <37  & <337  &  < 22 & < 197 &  -   & 218.7\,(8.6)    & 218.7\,(8.6) \\
G35.39$-$00.33 mm2  & A &  43 &  99 & < 45 & < 103  & 62.0   & 104.2\,(5.2)    & 120.3\,(10.0) \\
IRAS 18553+0414 & A1 & 65 &  802  &   < 34 & < 421  & 74.9   & 162.8\,(7.8)    & 201.9\,(15.6) \\
           &  A2 & 78 & 956   &   52  &  640   & 94.0   & 187.4\,(7.6)    & 292.3\,(17.9) \\
IRAS 18566+0408 &  A1 & 75 & 503   &   < 42 & < 283  & 62.3   & 64.3\,(3.9)     & 86.1\,(8.2) \\
            & A2 & 126 & 847 & < 43   & < 289  &   129.6 & 35.6\,(3.8) & 57.3\,(9.2) \\
            &  B & 53 &  356  &   27  & 182    & 142.7  & 133.2\,(3.9)    & 159.3\,(7.7) \\
IRAS 19012+0536 & A &  22 &  94  & 12 &  52  & 155.7  & 524.7\,(9.4)    & 543.2\,(16.7) \\
IRAS 19035+0641 & A2 & 57  & 131  &   < 21 & < 49   & 45.6   & 235.2\,(13.4)   & 276.6\,(25.9) \\
             & A3 & 55 &  126  &   < 11  & < 26  & 106.6  & 150.1\,(13.5)   & 165.3\,(24.9) \\
G53.11+00.05 mm2 &  A1 &80 & 151   &   < 9 & < 17   & 9.3    & 35.3\,(4.8)     & 42.1\,(9.3) \\
             &  A2 & 65 & 123  &   28 &  53  & 120.3  & 41.1\,(4.7)     & 57.3\,(10.2) \\
G53.25+00.04 mm2 & A & 67 & 127 & 28 & 53 & 59.0 & 111.2\,(7.1) & 156.0\,(16.0)\\
G53.25+00.04 mm4    &   A   &   <43    &  <86 &   <14    & <27  &  - &  39.1\,(5.0) & 39.1\,(5.0)  \\
IRAS 20293+3952 &  E & < 55   & < 71/109 & < 35 & < 46/71  & -  & 52.1\,(6.5)  & 52.1\,(6.5) \\ 
IRAS 20343+4129  &   B   &   76    & 107  &   52    & 72  &   160.1   &   42.0\,(6.4) &   64.0\,(14.8)\\
\enddata
\tablenotetext{}{The sizes given are deconvolved from the synthesized beam, and the values in parentheses are formal errors from the 2-D Gaussian fits. 
We excluded IRAS 19413+2332 due to its non-Gaussian behavior.}
\end{deluxetable*}


\section{Discussion}
\label{sec:discussion}
A radio jet is classically defined as a ``\textit{radio continuum source tracing a collimated ionized outflow from a YSO}'' \citep{Anglada1996}. 
Therefore, in order to unequivocally classify a radio source as a jet one would need to know the position of the driving protostellar object and be able to observe collimated ionized emission receding from this position.
Locating the central object is extremely challenging due their deeply embedded nature and large distances. 
Secondary probes can be used to obtain a constraint on the location of the MYSO, such as Class\,II CH$_3$OH masers.
These masers are pumped by intense infrared radiation and are exclusively detected in close proximity to massive protostars \citep{Minier2003,Cragg2005,Ellingsen2006, Moscadelli2011}.
We found detections of Class\,II CH$_3$OH masers at 6.7 GHz with a positional accuracy comparable to our synthesized beam size or better toward 11 sources in our survey: G11.11$-$0.12P1 (Figure \ref{fig:appendix-A2}), IRAS 18151$-$1208, IRAS 18182$-$1433 (Figure \ref{fig:appendix-A3}), IRAS 18264$-$1152 (Figure \ref{fig:appendix-A4}), G23.01$-$00.41, IRAS 18345$-$0641 (Figure \ref{fig:appendix-A5}), IRAS 18517+0437 (Figure \ref{fig:appendix-A6}), IRAS 18566+0408, IRAS 19012+0536, and IRAS 19035+0641 (Figure \ref{fig:appendix-A8}) \citep{Bartkiewicz09,Sanna10,Sanna10b,Araya10, Pandian11, Surcis15, Rosero2015}. 
These are marked with red circles in Figure \ref{fig:examples} and in the respective \hyperref[sec:appendix]{Appendix} images.
However, Class\,II CH$_3$OH masers are generally associated with accretion disks \citep[e.g.,][]{Burns2023}, which can have radii of 1,000s au as aforementioned.
Thus, their detection at close projected distance from the sources in our sample indicates the radio continuum emission is located in the vicinity of the MYSO, but the position of the protostar is still effectively unknown.

Being able to observe collimated ionized emission arising from a massive protostellar jet is also a strenuous task, as the thermal jet brightness is inversely proportional to the distance to the protostar squared and their angular size decreases with frequency \citep{Reynolds1986,Anglada2018}.
This, together with the large distances at which MYSOs are typically found, result in radio jets being usually unresolved. 
Hence, despite being seemingly associated with molecular or atomic outflows or other jet/shock tracers, most of the single compact sources in our sample
cannot be unambiguously confirmed as ionized jets, thus, the following sources remain as jet candidates: UYSO1 A (Figure \ref{fig:appendix-A1}), IRAS 18345$−$0641 A (Figure \ref{fig:appendix-A5}), IRAS 18470$-$0044, IRAS 18517+0437 A (Figure \ref{fig:appendix-A6}), IRAS 18521+0134 A, G35.39$−$00.33 mm2 A (Figure \ref{fig:appendix-A7}), IRAS 19012+0536 A (Figure \ref{fig:appendix-A8}), G53.11+00.05 mm2 A, G53.25+00.04 mm2 A, G53.25+00.04 mm4 A (Figure \ref{fig:appendix-A9}), IRAS 19413+2332, IRAS 20293+3952 E, and IRAS 20343+4129 B (Figure \ref{fig:appendix-A10}).
Based on their multi-peaked and elongated morphology, IRAS 18264$−$1152 F (Figure \ref{fig:appendix-A4}), G34.43+00.24 mm1 A (Figure \ref{fig:appendix-A6}), and IRAS 18553+0414 A (Figure \ref{fig:appendix-A7}) are strong ionized jet candidates. 
However, we cannot discard the possibility of the radio continuum emission probing individual YSOs, especially for IRAS 18264$−$1152, where multiple dust cores have been reported \citep{Gieser2023}, and G34.43+00.24 mm1, which is associated with a quadrupolar molecular outflow \citep{Isequilla21}.
For consistency, we will keep the ionized jet denomination for the following sources classified as such by \cite{Rosero19}: G11.11$−$0.12P1 (Figure \ref{fig:appendix-A2}), IRAS 18151$−$1208, IRAS 18182$−$1433 (Figure \ref{fig:appendix-A3}), G23.01$−$0.41, IRAS 18440$−$0148 A (Figure \ref{fig:appendix-A5}), IRAS 18566+0408, and IRAS 19035+0641 A (Figure \ref{fig:appendix-A8}).
We note that our higher angular resolution observations did not recover the extended features that granted them this classification seen toward IRAS 18440$-$0148 and IRAS 18151$−$1208 by \cite{Rosero16,Rosero19}, although we detected proper motion in both cases (see Section \ref{sec:discussion-PM}). While we try to keep consistency between surveys in terms of classification, these two cases should be treated carefully.  
In summary, our observations are consistent with the ionized jet classification of 7 sources, and further strengthen the ionized jet classification of 16 candidates.

\subsection{Comments on the compact morphologies observed}
\label{sec:morphology} 
Almost half the jets and jet candidates in this survey present a mostly compact appearance of size $\sim100-1,000\,$au, similarly to other sub-arcsecond resolution radio continuum studies \citep[e.g.,][]{Moscadelli2016,Kavak2021,Purser2021}.
As aforementioned, we generally recover less emission than \cite{Rosero16}, possibly due to the lack of shorter baselines in the array configuration we used for the higher angular resolution observations.
If the origin of the ionization is indeed shock ionization, 
this suggests the presence of core/halo shock structures.
That is, the radio continuum emission detected in our higher angular resolution observations probes the hot, compressed gas in the shock front (``core''). 
This gas radiates and ionizes the less compressed gas in the vicinity (``halo''), which is fainter and more extended, hence not detected with our smaller synthesized beam.
The fact that the string-like jets in the \cite{Rosero16} survey are mostly seen as multiple well-aligned, mostly compact radio continuum peaks in our data is consistent with this scenario.
Furthermore, the measured apparent deconvolved sizes suggest the core/halo structure exists on even smaller scales.

\subsection{Radio continuum proper motion}
\label{sec:discussion-PM}
Approximately a decade separates our observations and those of \cite{Rosero16,Rosero19}\footnote{Their VLA \textit{K}-band survey observations were carried out in the 2010B semester, except in the case of UYSO1, G53.11+00.05 mm2, G53.25+00.04 mm2, IRAS 19413+2332, IRAS 20343+4129, and IRAS 20293+3952, which were observed in the 2014A semester due to scheduling constrains.}, hence we investigated whether proper motion is detected.
We measured the projected separation of the 1.3 cm continuum peak position between the two epochs, as well as the projected velocity, and the proper motion of the jets and jet candidates that present a similar morphology in both observations, that is, that are a single, compact source.
Our results are given in Table \ref{tab:PM}.  
The peak positions utilized for these calculations were obtained via 2-D Gaussian fits. 
For the error estimation, we quantified the relative positional uncertainty as described in Equation 1 in \cite{Goddi2004} and a phase calibration error of 10\,mas, which is adequate for the VLA in the A-configuration at 1.3\,cm and appropriate since the same calibrators were used in both observing epochs.
We note, however, that there are other uncertainties at play that cannot be quantified, such as the signal path differences between the calibrator and the target.

Considering only those cases in which the peak separation is at least twice the uncertainty, we detected proper motion in 10 cases.
The measured separation in the sky in these cases range between 41.4\,mas (IRAS 18517+0437 A) and 105.6\,mas (G35.39$-$00.33 mm2), with an average of $\sim65\,$mas. In linear scales, the measured separations range between 77.0\,au (IRAS 20343+4129 B) and 781.3\,au (IRAS 18521+0134 A).
We measured proper motion velocities between $\sim34\,$\kms\ (IRAS 18517+0437 A) and $\sim337\,$\kms\ (IRAS 18521+0134 A), with an average of $\sim120\,$\kms.
As shown in Table \ref{tab:PM}, UYSO1, G23.01$-$00.41, G53.25+00.04, IRAS 20293+3952, and IRAS 20343+4129 were observed in 2013--2014. Hence, for completion, we also measured the separation, velocity and proper motion between their 1.3\,cm and 6\,cm continuum peaks, as the latter were all carried out in 2011. The values are consistent within errors with those we report, except in the case IRAS 20293+3952 E, which is not detected at 6\,cm.

The measured plane-of-sky velocities are on the lower end of what is usually observed in spatially resolved high-mass radio jets \citep[see][and references therein]{Anglada2018}.
However, we note that the viewing angle or inclination $i$ changes the position offset and velocity by a factor of $sin(i)$, and thus the reported values should be considered lower limits.
The relatively large proper motion and velocity values measured strongly support the interpretation of thermal jets being the origin of the ionized emission observed in the selected cases. 
This is especially true for the jet candidates with $V_{\mathrm{proj}}>100\,$\kms, namely IRAS 18440$-$0148 B, IRAS 18470$-$0044 B (Figure \ref{fig:appendix-A6}), IRAS 18521+0134 A, G35.39-00.32 mm2 A (Figure \ref{fig:appendix-A7}), and IRAS 19012+0536 A (Figure \ref{fig:appendix-A8}).
Note that among these, only IRAS 19012+0536 A has large-scale molecular outflow data available.
The last two columns of Table \ref{tab:PM} list, for each source, the directions of the measured proper motion and of the molecular outflow axis. 
A good alignment between these directions is found for the G23.01$-$00.41 A jet (Figure \ref{fig:appendix-A5}) and for the jet candidates IRAS 19012+0536 A (Figure \ref{fig:appendix-A8}) and IRAS 20343+4129 A (Figure \ref{fig:appendix-A10}), further supporting the interpretation of these sources as ionized jets.

\begin{deluxetable*}{l c c c c c c c c}[htpb]
\tablenum{5}
\label{tab:PM}
\tabletypesize{\scriptsize}
\tablecaption{Proper motion (PM) of selected sources.}
\tablecolumns{7}
\tablewidth{0pt}
\tablehead{
\colhead{Source} & \colhead{Obs. Date 1} & \colhead{Obs. Date 2} & \multicolumn{2}{c}{$\Delta$position$_\mathrm{proj}$} & \colhead{$V_\mathrm{proj}$} & \colhead{PM} & \multicolumn{2}{c}{Direction}\\
\colhead{} & \colhead{(dd/mm/yyyy)} & \colhead{(dd/mm/yyyy)} & \colhead{(mas)} & \colhead{(au)} & \colhead{(km s$^{-1}$)} &  \colhead{(mas yr$^{-1}$)} & \colhead{PM\hyperlink{note1}{\tablenotemark{$^a$}}} & \colhead{Outflow\hyperlink{note2}{\tablenotemark{$^b$}}}
}
\startdata
UYSO1 A	&	24/11/2013 &	14/05/2022 &	33.2 (18.7) &	33.2 (18.7) &	 18.6 (10.4) &	3.9 (2.2)  &  ...   & NW-SE  \\
UYSO1 B &   24/11/2013 &    14/05/2022 &    21.7 (39.9) &  21.7 (39.9) &     12.1 (22.3) & 2.6 (4.7)  & ... &  NW-SE \\
G11.11$-$0.12P1 A&	20/03/2011 & 25/04/2022 & 65.9 (58.9) &	189.1 (169.2) &	80.8 (72.2) &	5.9 (5.3) & ... &  E-W, NE-SW \\ 
G11.11$-$0.12P1 B&	20/03/2011 &	25/04/2022 &	95.4 (58.5) &	273.6 (167.9) &	116.9 (71.7) &	8.6 (5.3) & ... &  E-W, NE-SW \\ 
IRAS 18151$-$1208 B	&	20/03/2011 &	18/05/2022 & 51.0 (14.7)&	101.5 (29.3) &  43.1 (12.4) &	4.6 (1.3) & N-S & NW-SE  \\ 
G23.01$-$00.41 A&	18/01/2014 &	06/05/2022 & 45.0 (14.6) &	206.3 (67.0) &	117.9 (38.3) &	5.4 (1.8)  & NE-SW & NE-SW  \\ 
IRAS 18440$-$0148 A	&	02/05/2011 &	07/03/2022 &	70.3 (16.3) &	365.5 (84.6) &	159.7 (37.0) &	6.5 (1.5) & NE-SW &  - \\
IRAS 18470$-$0044 B	&	02/05/2011 &	09/04/2022 &	59.1 (23.8) &	385.7 (155.2) &	167.2 (67.3) &	5.4 (2.2) & NE-SW&  E-W \\ 
IRAS 18517+0437 A	&	14/04/2011 &	09/04/2022 &	 41.4 (16.4) &	78.6 (31.1) &  	33.9 (13.4) &	3.8 (1.5) & N-S & NW-SE  \\ 
IRAS 18521+0134 A	&	14/04/2011 &	09/04/2022 &	85.9 (16.4) &	781.3 (149.0) &	337.1 (64.3) &	7.8 (1.5) & NE-SW & -  \\ 
G35.39$-$00.33 mm2 A&	14/04/2011 &	27/03/2022 &	105.6 (22.2) &	243.0 (51.1) &	 105.2 (22.1) &	9.6 (2.0) &NE-SW & -  \\
IRAS 19012+0536 A	&	15/04/2011 &	30/04/2022 &	69.8 (14.3) &	293.2 (60.1) &	125.9 (25.8) &	6.3 (1.3) & NE-SW & NE-SW  \\ 
G53.25+00.04 mm2&	17/01/2014 &	05/05/2022 &	 70.7 (23.1) &	141.5 (46.2) &	80.9 (26.4) & 8.5 (2.8) &NE-SW &  - \\ 
G53.25+00.04 mm4&	17/01/2014 &	05/05/2022 &	 32.7 (25.8) &	65.4 (51.5) &	37.4 (29.4) &	3.9 (3.1)  & ... &  - \\ 
IRAS 20293+3952 E &	18/01/2014 &	06/06/2022 &41.0 (23.2)\hyperlink{note3}{\tablenotemark{$^c$}} &	53.3 (46.4) &	30.2 (26.2) &	4.9 (2.8)  & ... & NW-SE  \\ 
IRAS 20343+4129 B	&	18/01/2014 &	06/06/2022 &	55.0 (21.8) &	77.0 (30.6) &	 43.5 (17.3) &	6.6 (2.6) & E-W & E-W  \\ 
\enddata
\tablenotetext{a}{\hypertarget{note1}{}Proper motion is considered detected when the displacement is at least twice the estimated uncertainty. No proper motion direction is assigned to non-detections.}
\tablenotetext{b}{\hypertarget{note2}{}See Table \ref{tab:cont-sources} for references.}
\tablenotetext{c}{\hypertarget{note3}{}Near distance assumed.}
\end{deluxetable*}

\subsection{Radio continuum variability}
\label{sec:variability}
The time span between the two observing epochs also provides the opportunity to investigate if brightness variability is detected \citep[e.g.,][]{Rodriguez2008,Carrasco-gonzalez2012_variability}. 
While a decrease in the flux density recovered in our observations in comparison with those of \cite{Rosero16} might be indicating a core/halo structure (see Section \ref{sec:morphology}), an increase in the measured flux density cannot be explained by this scenario.  
A significant brightness increase was detected in a few cases, one remarkable case being G11.11$-$0.12P1\,A$-$E. 
This string-like jet is composed of individual, well aligned radio continuum sources, three of which were detected at 1.3\,cm by \cite{Rosero16}: peaks A, B, and C. 
We find that the intensity of peak A has almost doubled (from 40 to 79\,\mujy) and that of peak B has decreased by half (from 75 to 37\,\mujy), while peak C, with an intensity of $50\,\mu$\Jyb\ in the \cite{Rosero16} data, was not detected in our observations (i.e., $I_{1.3cm}<15\,\mu$\Jyb).
In contrast, peak E, with an intensity of $34\,\mu$\Jyb\ in our 2022 observations, was not detected in the \cite{Rosero16} data. This suggests it was weaker than their $3\sigma$ detection limit of about $26\,\mu$\Jyb, indicating a brightness variation of $\gtrsim20\%$.
We interpret these discordant variations found as a combination of core/halo effect and radio variability. 

Another interesting case is that of G23.01$-$0.41\,A. 
This is a known jet powered by a MYSO surrounded by a disk-like structure undergoing prominent infall motions \citep{Sanna2019,Sanna2021}. 
\cite{Sanna10} observed in 2008 the 1.3\,cm continuum emission toward this jet with the VLA in the A-configuration (i.e., the same angular resolution as our observations)
and measured a peak intensity of $\sim720\,\mu$\Jyb, consistent with the $\sim730\,\mu$\Jyb\ peak emission measured by \cite{Rosero16}. 
However, we measured a peak intensity of $\sim370\,\mu$\Jyb, showcasing a brightness variability of approximately 50\% in 14 years.
Moreover, the integrated flux density measured by \cite{Sanna10} ($980\,\mu$Jy) also almost doubles ours ($511\,\mu$Jy). 
In addition, \cite{Sanna2016} measured peak and integrated intensities of 678\,\mujy\ and $775\,\mu$Jy, respectively, in their 2016 VLA B-configuration observations ($\theta\sim0$\farcs33).
These are values in between those measured in the 2008 and 2022 data. 
Based on the above, we discard a core/halo scenario or other observational effects as the cause of the variations found.

Even though the cause of short-term ($\sim1-10$\,yr) variability in jets is not well understood, it could be associated with accretion variability in high-mass star formation. 
Indeed, a similar behavior to that of G23.01$-$00.41\,A was observed in the jet propelled by the MYSO S255IR-NIRS3 after an accretion outburst \citep[e.g.,][]{Cesaroni2018,Cesaroni2023}.
Recent numerical simulations suggest that accretion variability, which is widely observed in low-mass star formation \citep{Fischer2023}, might be key in regulating the formation of MYSOs as well \citep[e.g.,][]{Klassen2016,Meyer2017,Oliva2023,Elbakyan2023}. 
Particularly, the magneto-hydrodynamic simulations of \cite{Meyer2021} suggest a bimodality in the accretion history of MYSOs, with low frequency ($\sim10^3-10^4\,$yr) intense outbursts, and high frequency ($\sim1-10\,$yr) weaker and fainter bursts. 
The former would lead to 
a brightness increase of a few factors to a few orders of magnitude across the entire spectrum \citep[e.g.,][]{Fujisawa2015,Hunter2017,Brogan2019, Proven-Adzri2019, Chen2021}, while the latter could explain the formation of internal jet knots but would have a more subtle effect in their surroundings.
After searching in the maser database\footnote{\url{https://www.maserdb.net}} 
run by the maser monitoring organization\footnote{\url{https://www.masermonitoring.com}}, 
as well as the Ibaraki 6.7\,GHz class II methanol maser database\footnote{\url{https://vlbi.sci.ibaraki.ac.jp/iMet/}}, 
we found that maser variability and relatively weak flares have been observed, but no intense maser activity has been recorded in the last decade toward any of the sources in our survey. 
Therefore, we consider that weak accretion/ejection bursts might be behind the observed radio jet variability.
We note that numerical simulations are able to model variable accretion with great precision now, but ionized condensations due to episodic accretion bursts have been proposed in high-mass star forming regions decades ago \citep[e.g., HH 80/81,][]{Marti1995}.

\section{Summary \& Conclusions}
\label{sec:summary}
We have carried out 1.3\,cm continuum and 22.2\,GHz H$_2$O maser observations toward all the jet candidates from the \cite{Rosero16,Rosero19} survey, together with 10 other sources to enrich the sample.
With $3\times$ higher angular resolution ($\theta\sim0$\farcs1), our observations were used to study the morphology of the jet candidates and search for elongated radio continuum structures, which would help confirm their jet nature. 
Our results can be summarized as follows:
\begin{enumerate}
    \item We detected 22 out of the 23 target sources and report on 1 additional jet candidate that had not been included in the \cite{Rosero16,Rosero19} study. We observe three types of morphologies: string-like, elongated and/or double-peak connected by low level continuum emission (6 out of 23 cases, 26\%), 2 or more individual (i.e., well spatially separated) continuum peaks (4 out of 23 cases, 17\%), and compact (13 out of 23 cases, 57\%).
    
    \item The majority of H$_2$O maser spots detected are within a sky-projected distance of $\sim5,000\,$au from the radio continuum peaks and are 
    well aligned and distributed in an elongated structure in some notable cases: IRAS 18182$-$1433, G23.01$-$00.41, G35.39$-$00.33 mm2, IRAS 18566+0408, IRAS 19012+0536, and IRAS 19035+0641.
    
    \item Based on their elongated or string-like morphology and/or proper motion detection, as well as the previous classification by \cite{Rosero19}, our observations are consistent with the ionized jet classification of 7 sources, and further strengthen the ionized jet classification of 16 candidates.
    
    \item We generally recover less emission than \cite{Rosero16}, which suggests what we observe are core/halo structures, that is, a very small shock front (``core'') that is surrounded by shock-ionized, less compressed gas (``halo''). In this sense, our observations are not sensitive to more extended emission from the fainter, less compressed gas.
    
    \item We detected proper motion toward 10 single, compact radio continuum sources in the survey. The average projected velocity measured is $\sim120\,$\kms. 
    The viewing angle is unknown, thus the proper motion velocity values should be taken as a lower limit to the total velocity. 
    The proper motion direction coincides with the large scale molecular outflow axis in 3 out of the 6 cases in which the latter is known (namely, the G23.01$-$00.41 A jet and the jet candidates IRAS 19012+0536 A and IRAS 20343+4129 B), which provides additional support to the ionized jet nature of these candidates.
    
    \item Radio continuum variability was detected in a few sources, the most striking cases being the G11.11$-$0.12P1 and G23.01$-$0.41 thermal jets.
    \begin{itemize}
        \item The brightness of all the jet knots in G11.11$-$0.12P1 varied by $\sim20-70$\%, some increased while others decreased.
        \item Comparing ours with previous observations carried out with the VLA with the same angular resolution and at the same frequency \citep{Sanna10}, we found that the G23.01$-$00.41\,A jet is 50\% less bright than in 2008. 
    \end{itemize}
    We propose that the radio continuum variability detected in some of the thermal jets in our survey could be associated with weak accretion bursts. 
\end{enumerate} 
%
This work complements our previous search for molecular jet association as confirmation of the jet nature of the candidates \citep{Rodriguez23} and further characterizes ionized jets and jet candidates from \cite{Rosero16,Rosero19}.
The next generation of radio telescopes will allow more in depth studies of these radio jet candidates. With higher sensitivity to faint, extended emission and higher angular resolution, observations with instruments like the ngVLA should unveil the morphology of the radio emission on scales of $\lesssim100\,$au. 

\bigskip

\small
\noindent\textit{Acknowledgments.} We thank the anonymous reviewer for the warm reception of this manuscript and for their insightful comments and suggestions, which highly enriched this work.
Support for this work was provided by the NSF through
the Grote Reber Fellowship Program administered by Associated Universities, Inc./National Radio Astronomy Observatory. 
P. H. and E. D. A. acknowledge support from NSF grants AST–1814011, and AST–1814063, respectively. 
V. R. acknowledges support from NSF grant AST–2206437 to the Space Science Inst.
This work was partially funded by the NRAO's Student Observing Support SOSP22A--001.

\bibliography{biblio}{}
\bibliographystyle{aasjournal}

\appendix
\label{sec:appendix}
\section{Continuum Images}

The following images show the 1.3 cm continuum emission in our sample.
The red circles show the 6.7\,GHz CH$_3$OH masers reported by \cite{Bartkiewicz09} (IRAS\,18345$-$0641), \cite{Sanna10b} (IRAS\,18182$-$1433), \cite{Sanna10} (G23.01$-$0.41), \cite{Araya10} (IRAS\,18566+0408),  \cite{Pandian11} (IRAS\,19012+0536, IRAS\,19035+0641, G53.11+00.05\,mm2), \cite{Surcis15} (IRAS\,18517+0437), and \cite{Rosero2015} (IRAS\,18151$-$1208, IRAS\,18264$-$1152, and G11.11$-$0.12P1).
\renewcommand{\thefigure}{\Alph{section}.\arabic{figure}}
\setcounter{figure}{0}

\begin{figure}[!ht]
\centering
\centering
\includegraphics[width=.9\textwidth]{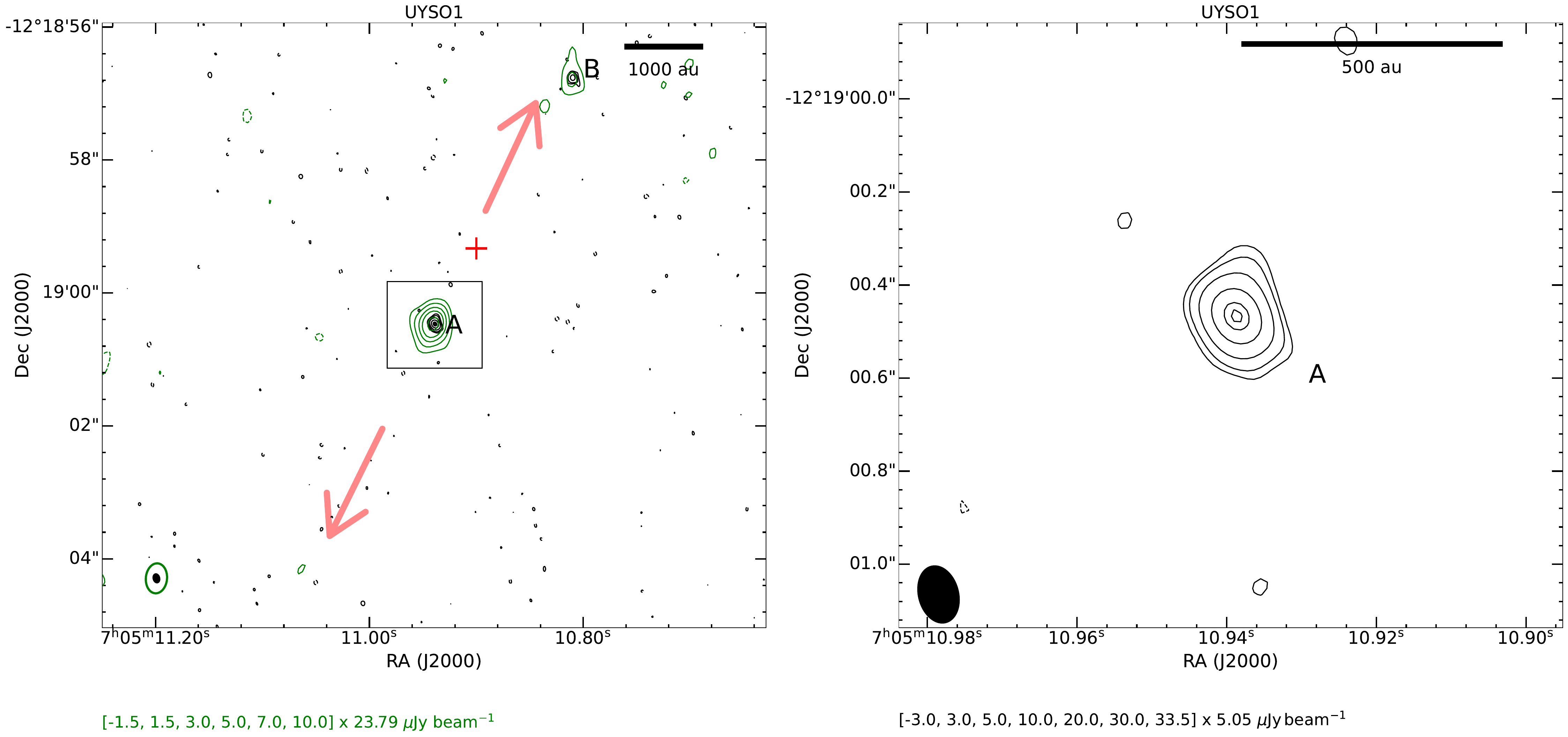}
\caption{VLA 1.3 cm continuum emission from \cite{Rosero16} and this work are shown in green and black contours, respectively. A zoom-in panel of our new, higher angular resolution data is included to showcase morphological details.
The black rectangle drawn on the larger area panel marks the area covered in the zoom-in panel. The pink arrows show the direction of the large scale molecular outflow axis as reported in the literature (see Table \ref{tab:cont-sources}) and the red + symbols show the \water\ masers detected in our observations.
The synthesized beam sizes are displayed with ellipses in the bottom left corner of each panel and a scale bar is shown in the top right. For those sources with a distance ambiguity, a near/far scale bar is given.}
\label{fig:appendix-A1}
\end{figure}

\begin{figure}[!ht]
\centering
\includegraphics[width=.4\textwidth]{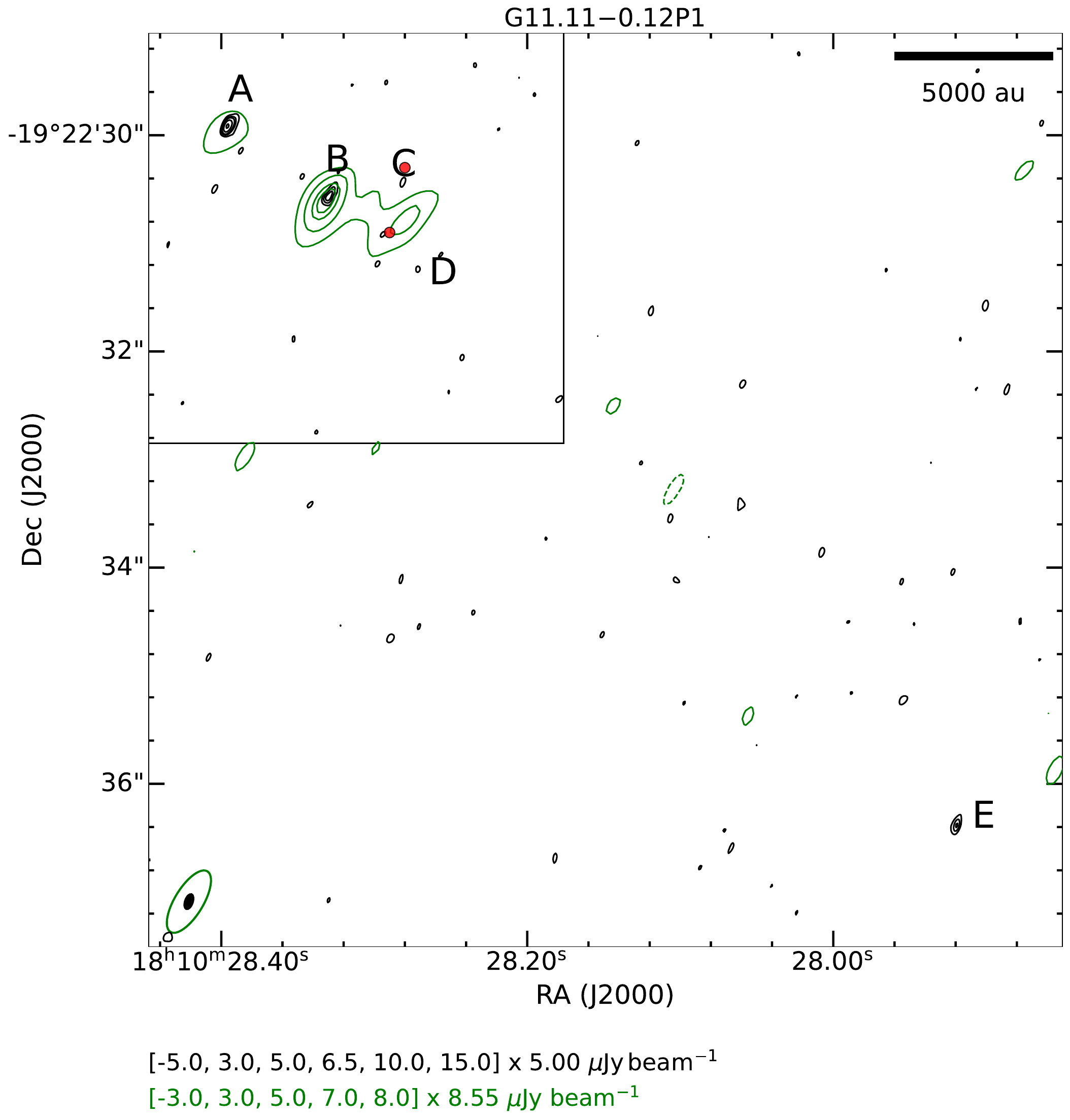}
\includegraphics[width=.9\textwidth]{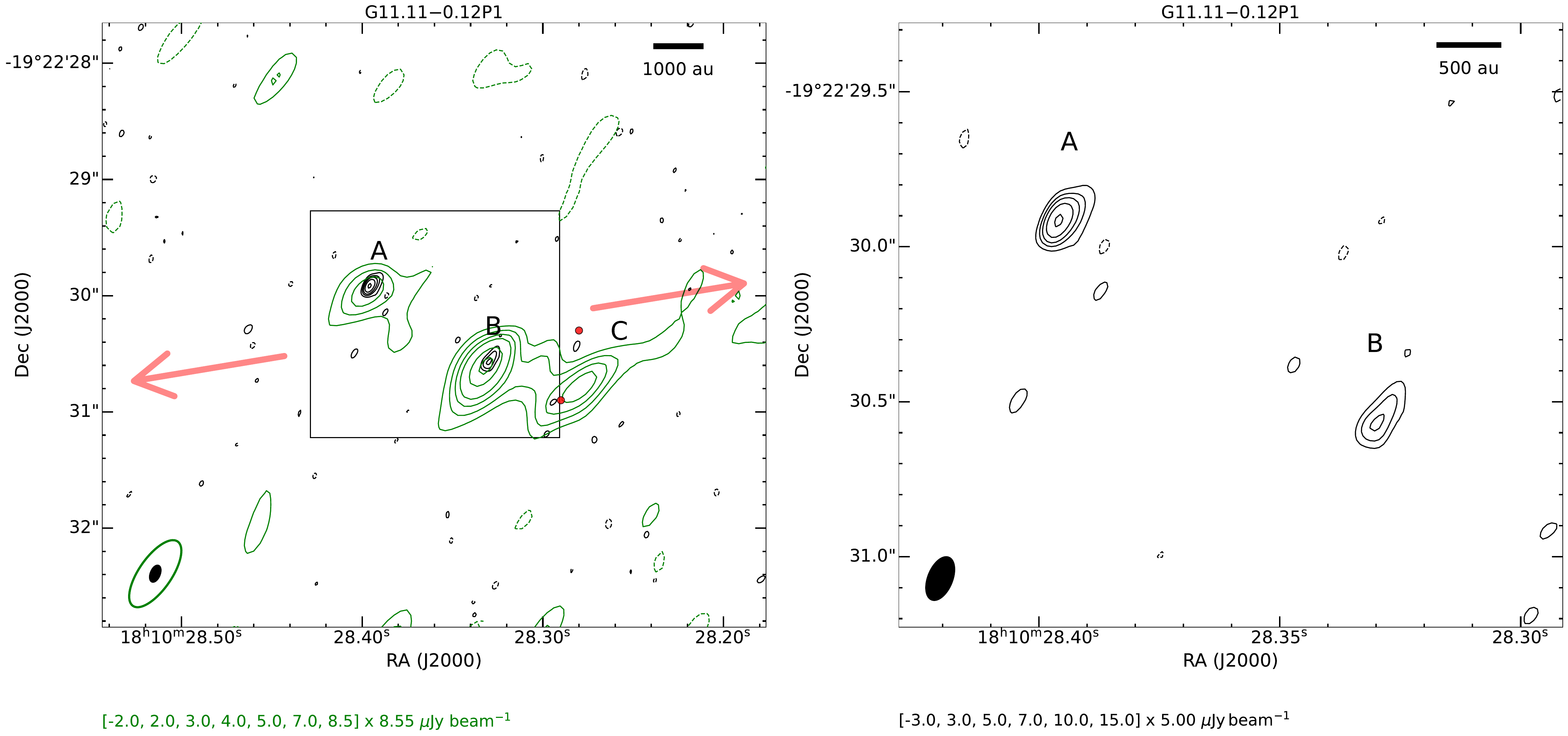}
\caption{Same as Figure \ref{fig:appendix-A1} but for G11.11$-$0.12P1. An additional larger panel was included to show nearby radio continuum features of interest.}
\label{fig:appendix-A2}
\end{figure}

\begin{figure}[!ht]
\centering
\includegraphics[width=.9\textwidth]{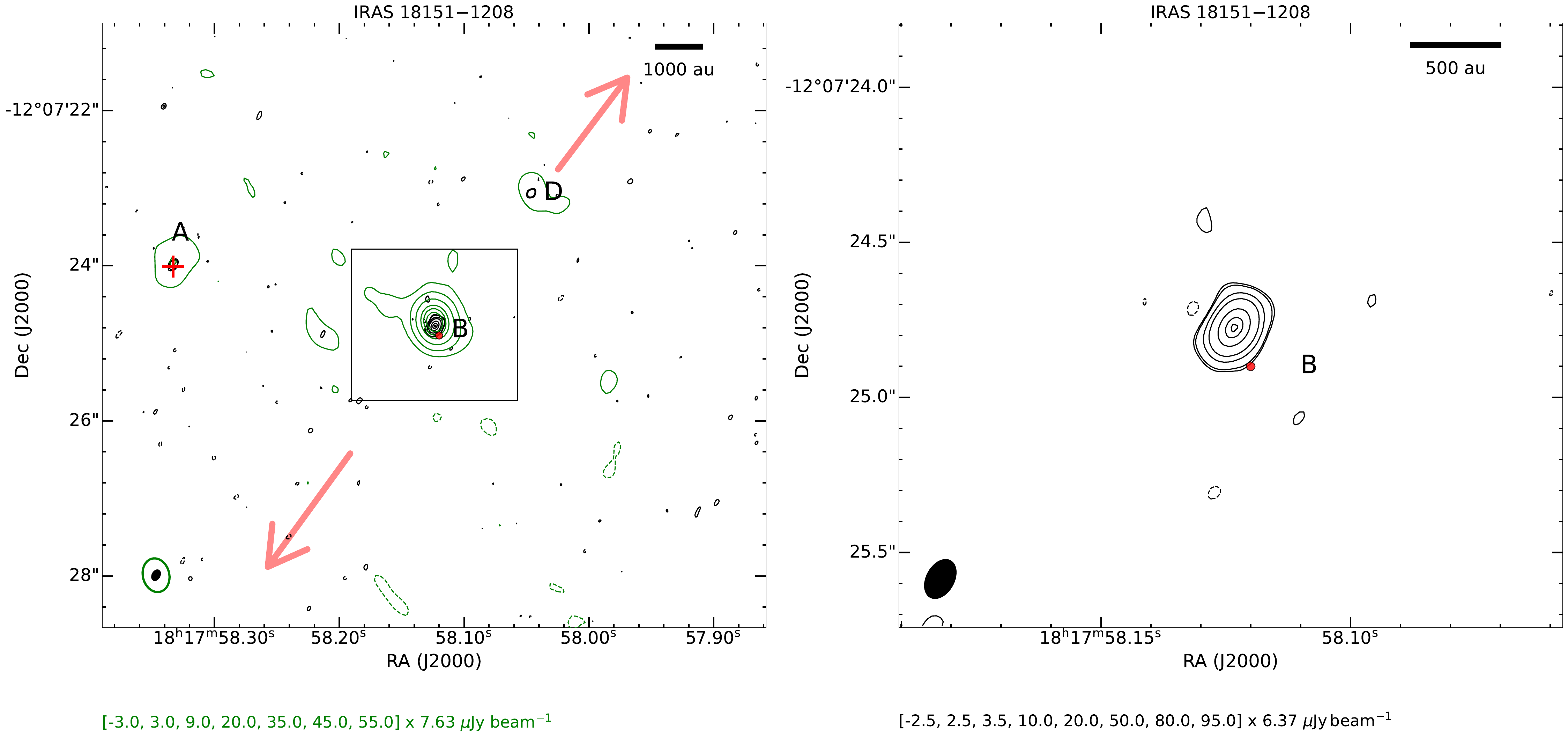}
\includegraphics[width=.9\textwidth]{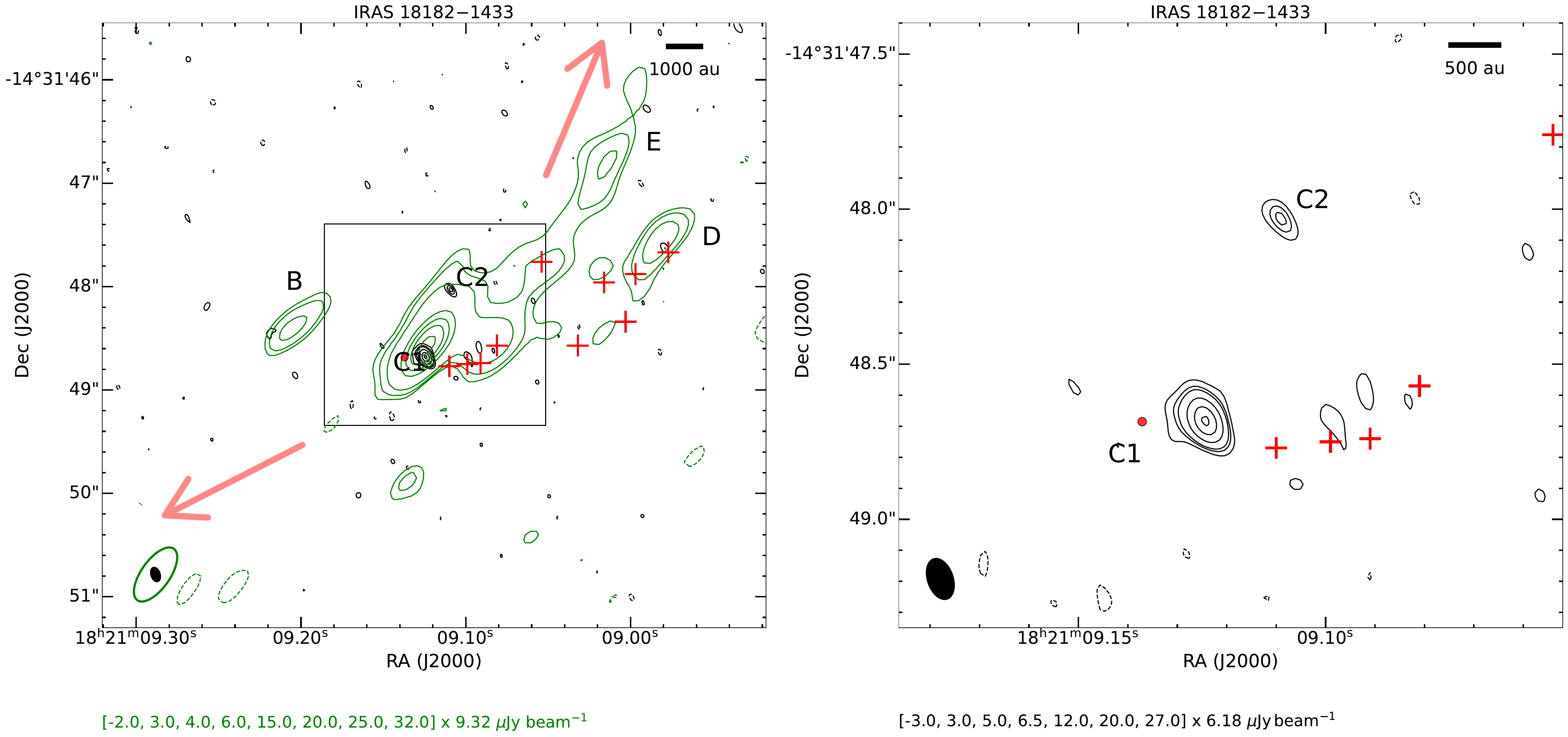}
\includegraphics[width=.42\textwidth]{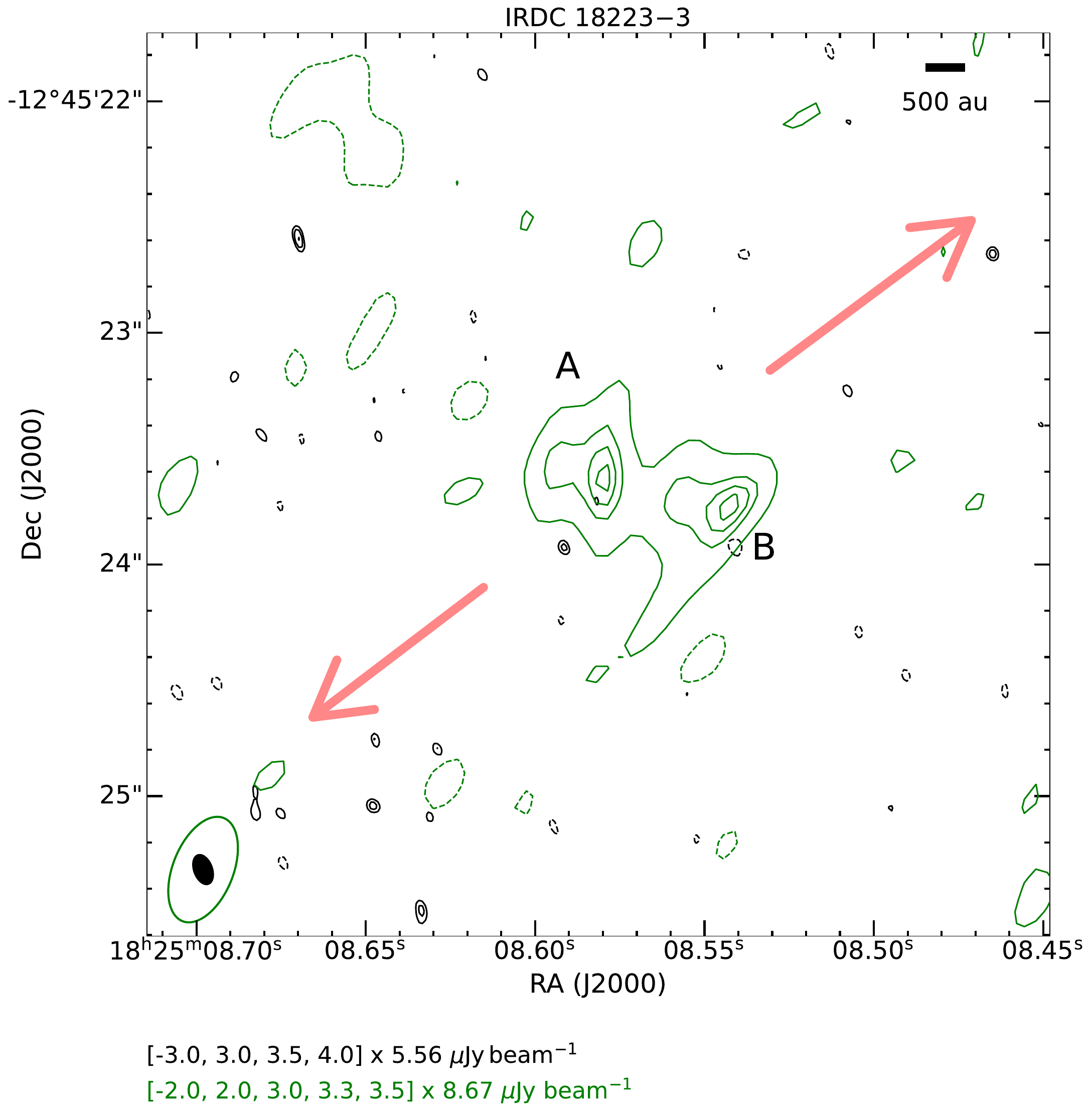}
\caption{Same as Figure \ref{fig:appendix-A1} but for IRAS 18151$-$1208 (top), IRAS 18182$-$1433 (middle), and IRDC 18223$-$3 (bottom). }
\label{fig:appendix-A3}
\end{figure}

\begin{figure}[!ht]
\centering
\includegraphics[width=.42\textwidth]{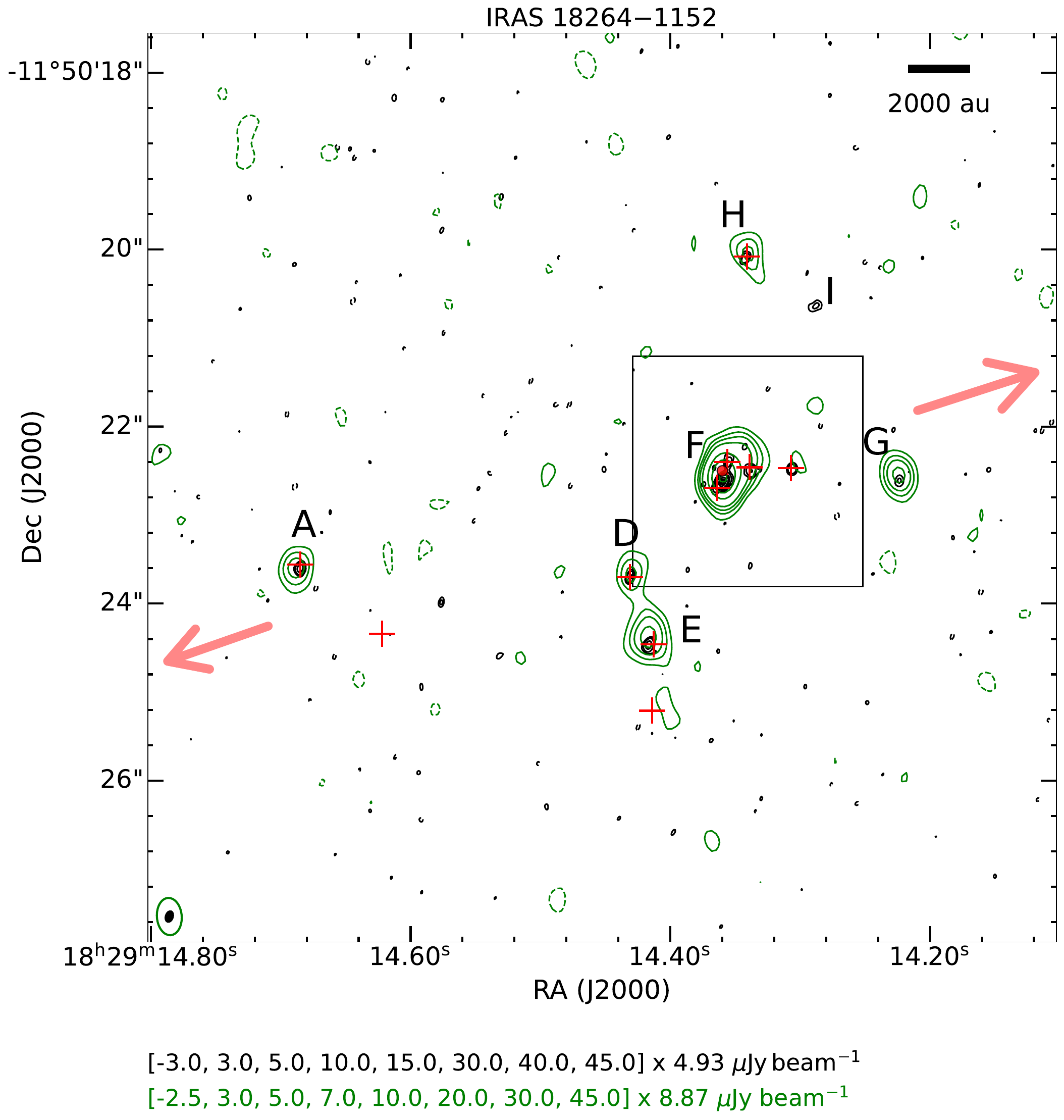}\\
\includegraphics[width=.9\textwidth]{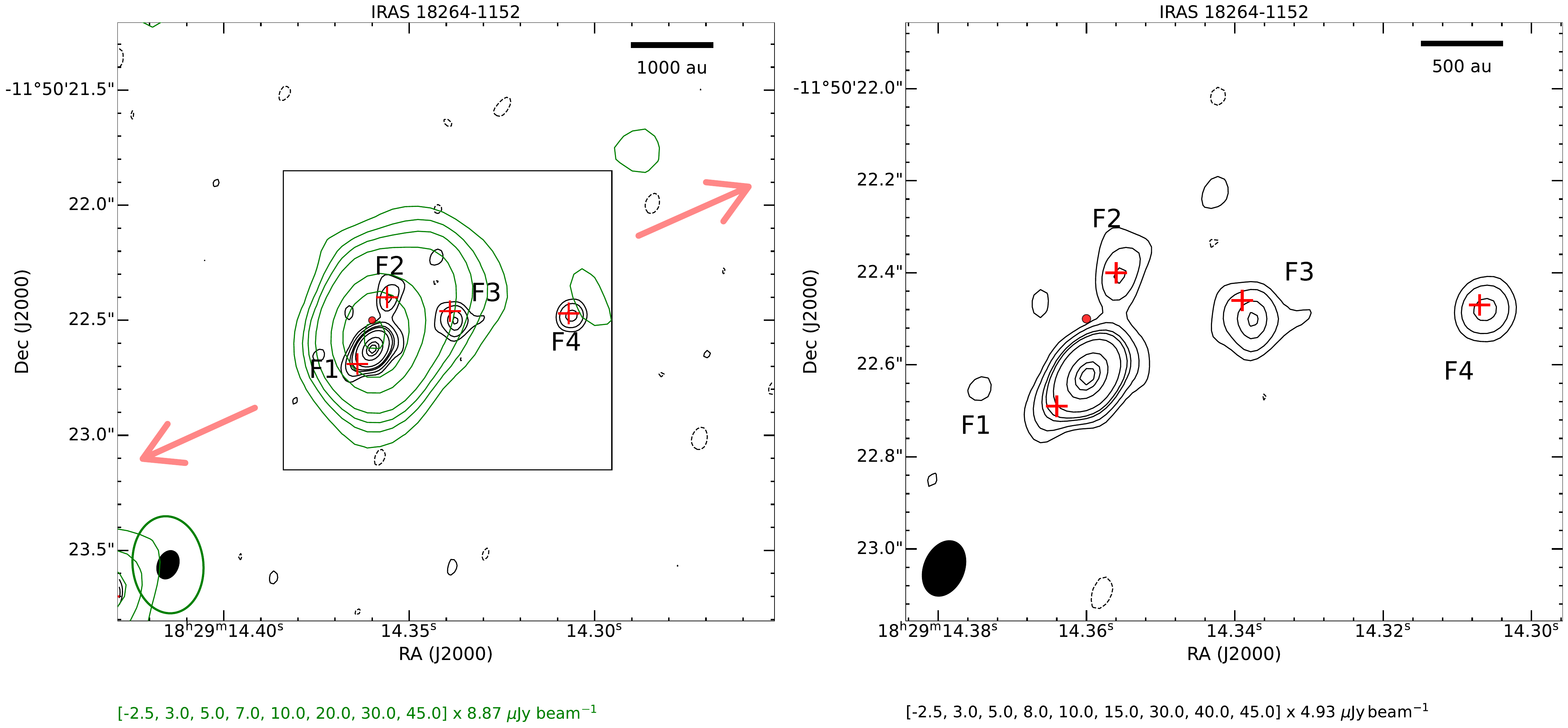}
\caption{Same as Figure \ref{fig:appendix-A1} but for IRAS 18264$-$1152. An additional larger panel was included to show nearby radio continuum features of interest.}
\label{fig:appendix-A4}
\end{figure}

\begin{figure}[!ht]
\centering
\includegraphics[width=.9\textwidth]{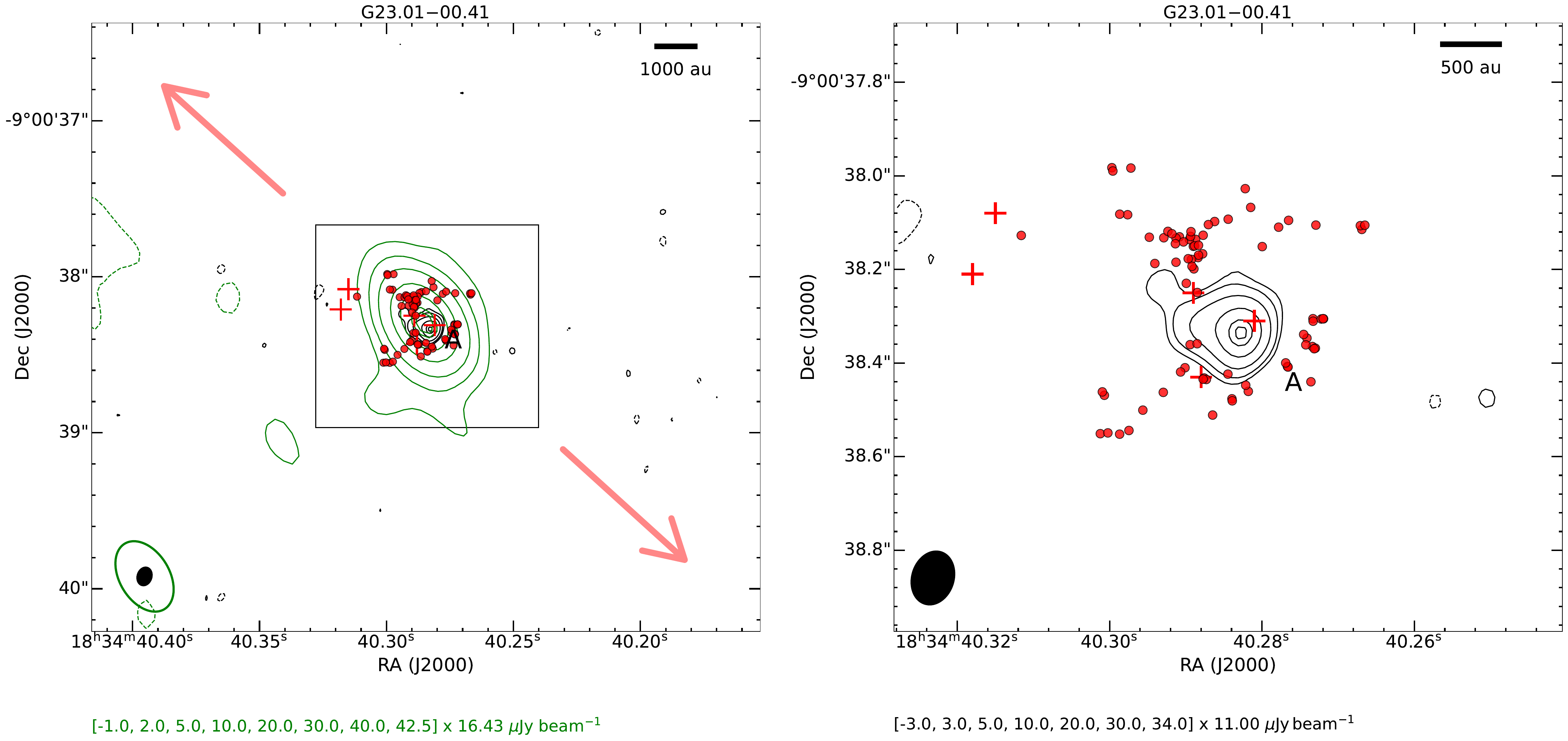}
\includegraphics[width=.9\textwidth]{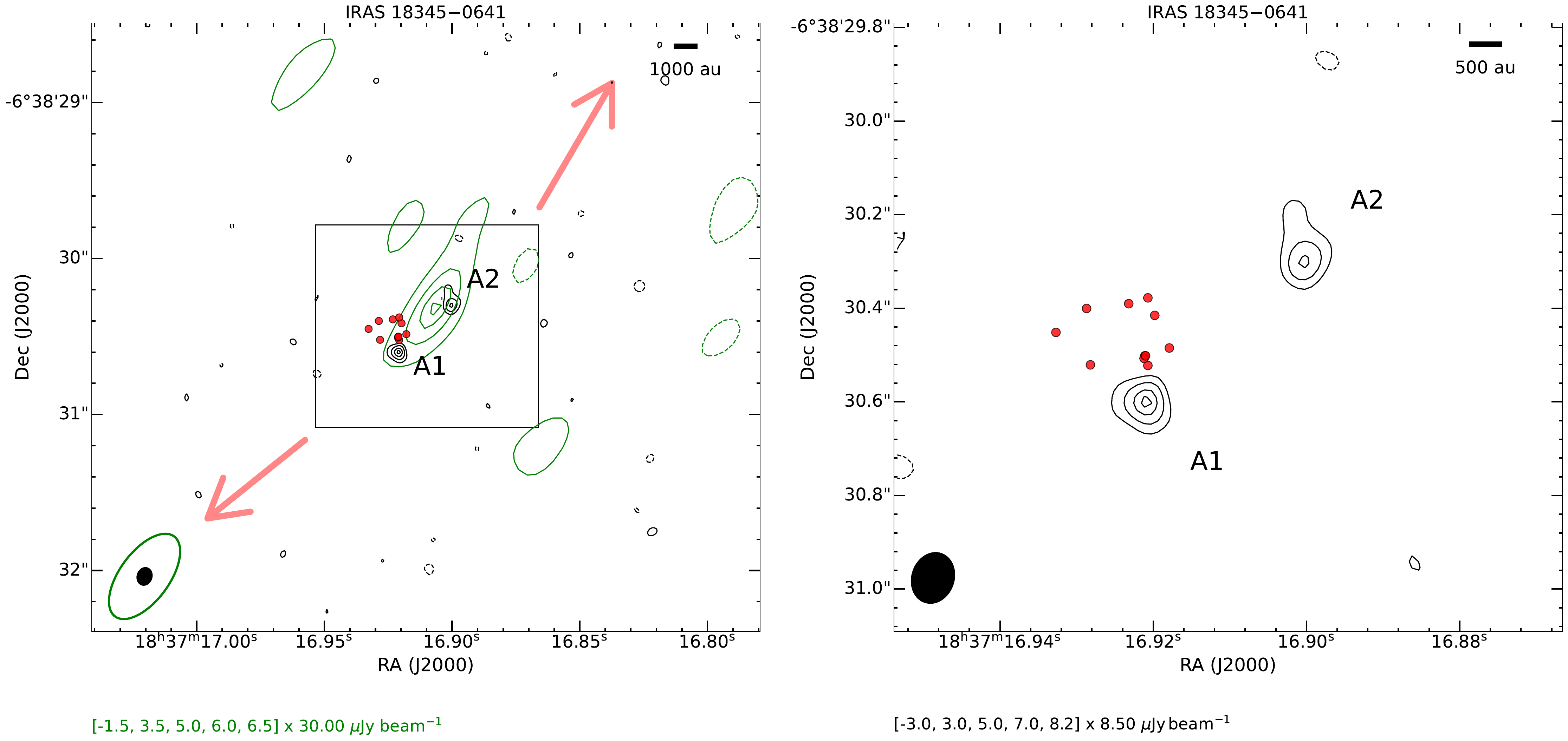}
\includegraphics[width=.9\textwidth]{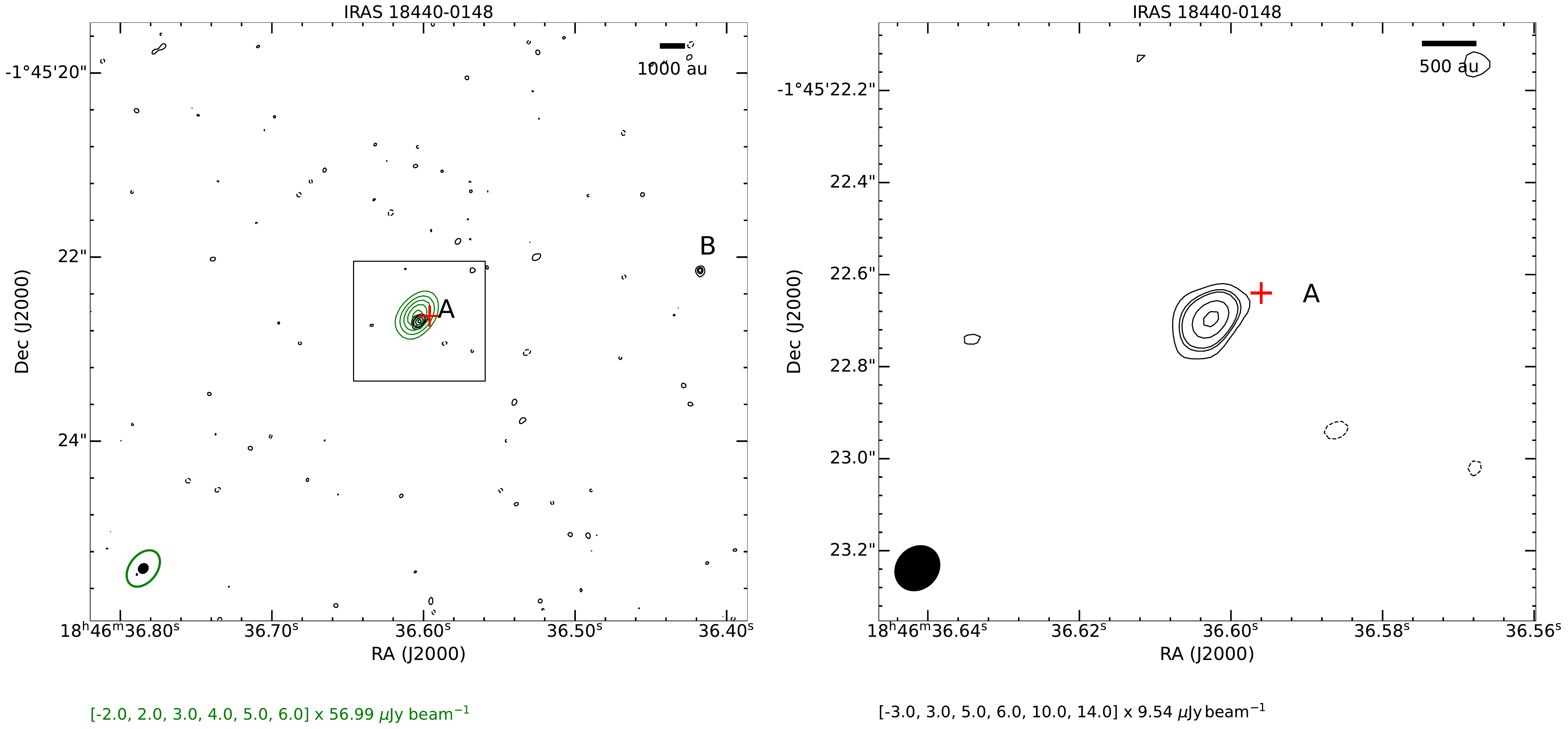}
\caption{Same as Figure \ref{fig:appendix-A1} but for G23.01$-$00.41 (top), IRAS 18345$-$0641 (middle), and IRAS 18440$-$0148 (bottom). }
\label{fig:appendix-A5}
\end{figure}

\begin{figure}[!ht]
\centering
\includegraphics[width=.9\textwidth]{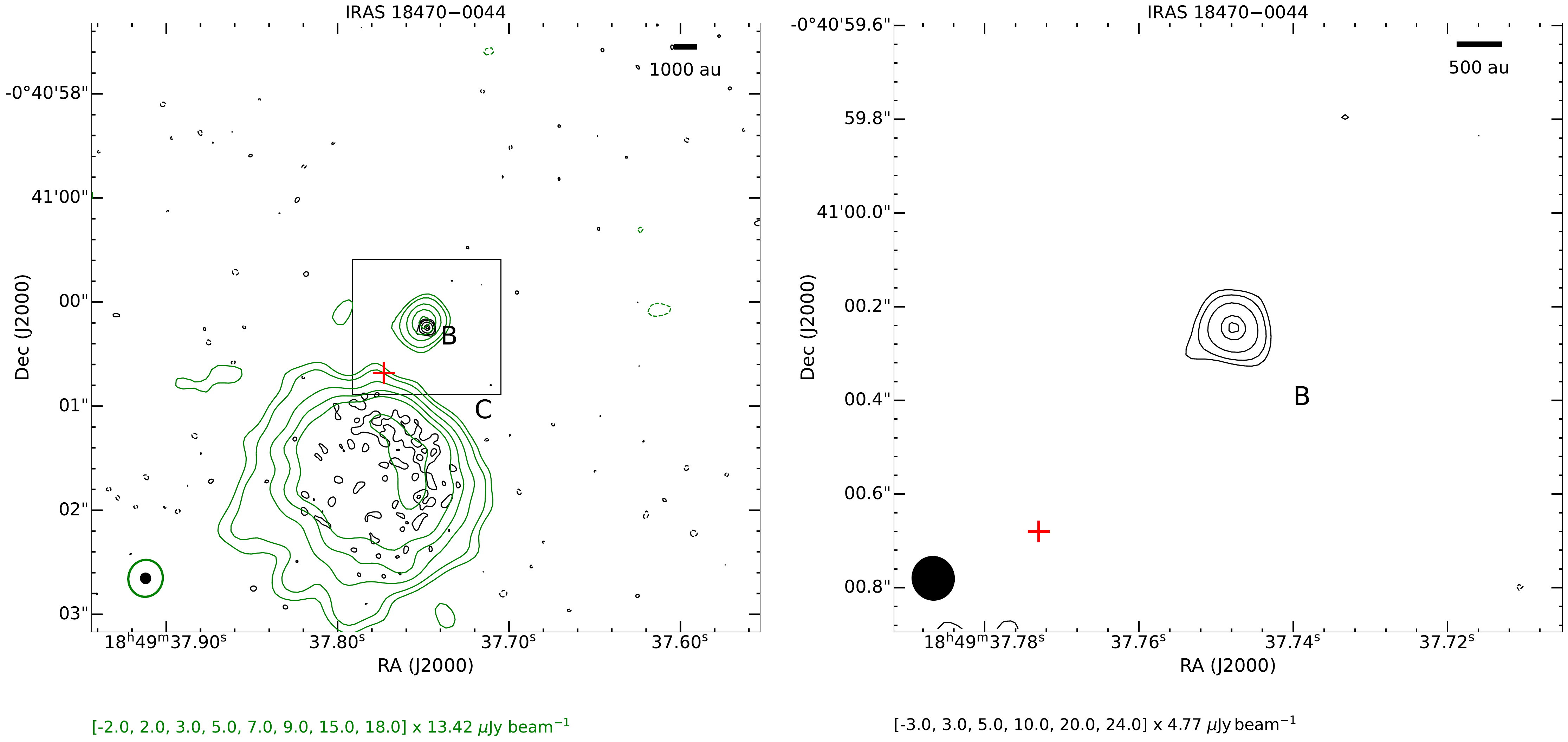}
\includegraphics[width=.9\textwidth]{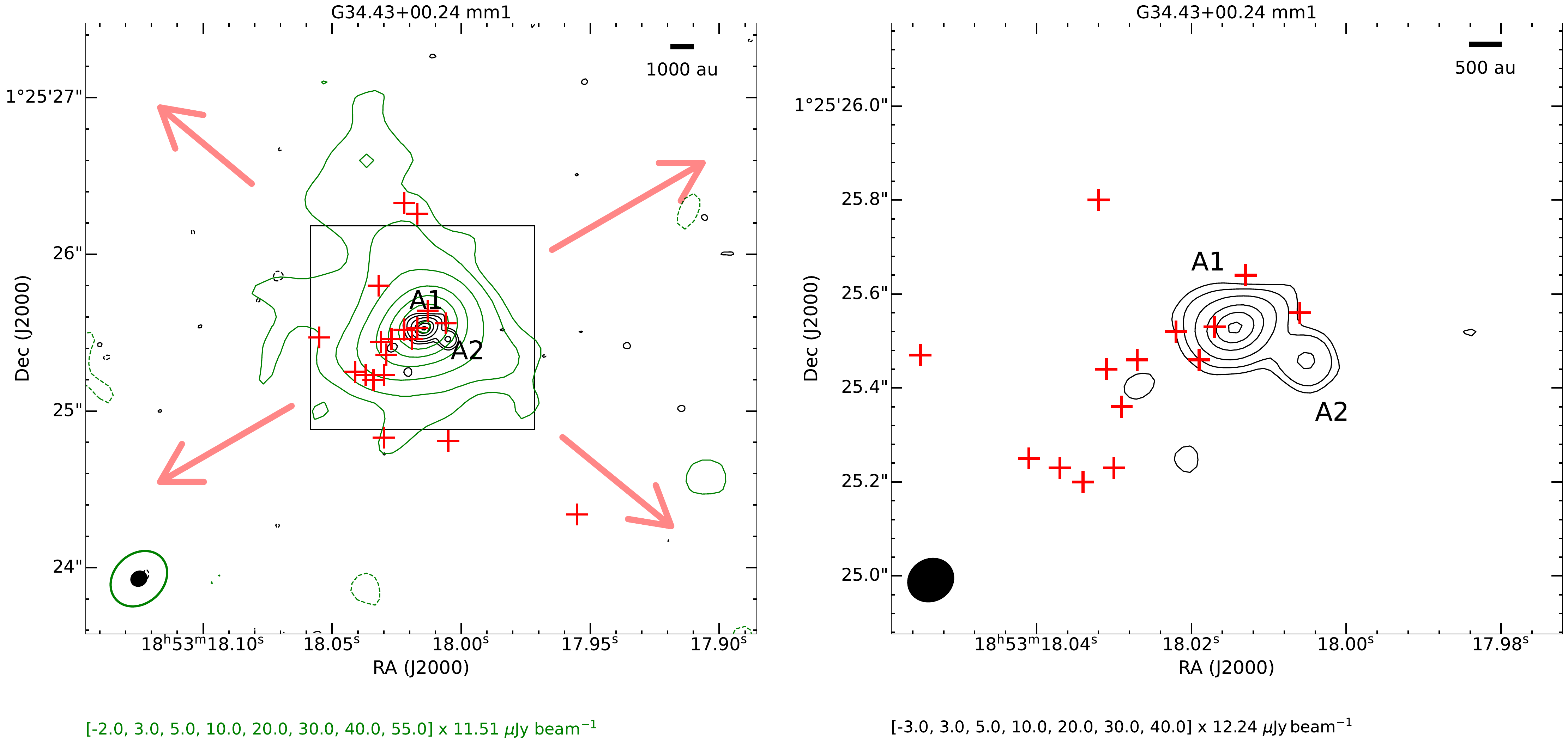}
\includegraphics[width=.9\textwidth]{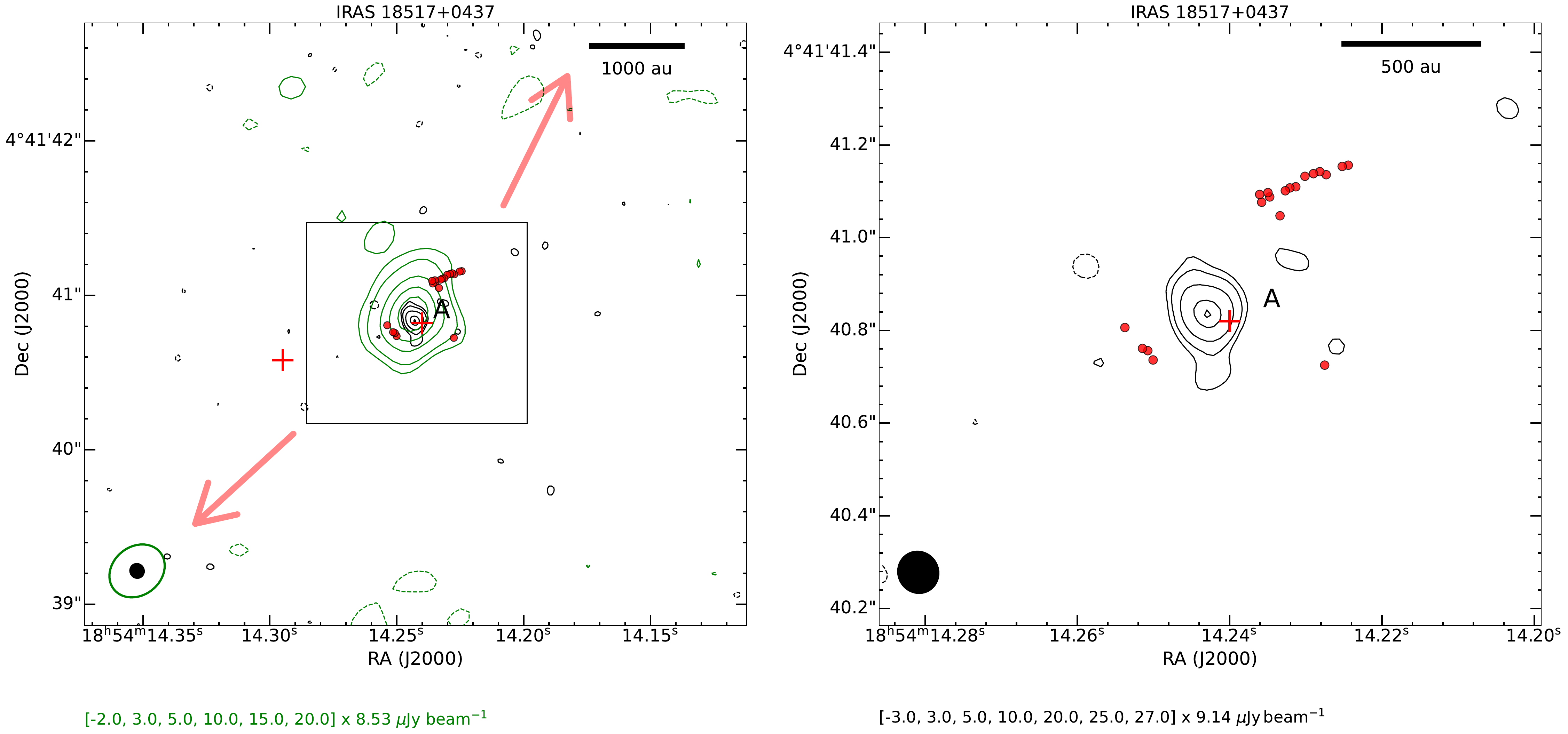}
\caption{Same as Figure \ref{fig:appendix-A1} but for IRAS 18470$-$0044 (top), G34.43+00.24 mm1 (middle), and IRAS 18517+0437 (bottom). }
\label{fig:appendix-A6}
\end{figure}

\begin{figure}[!ht]
\centering
\includegraphics[width=.9\textwidth]{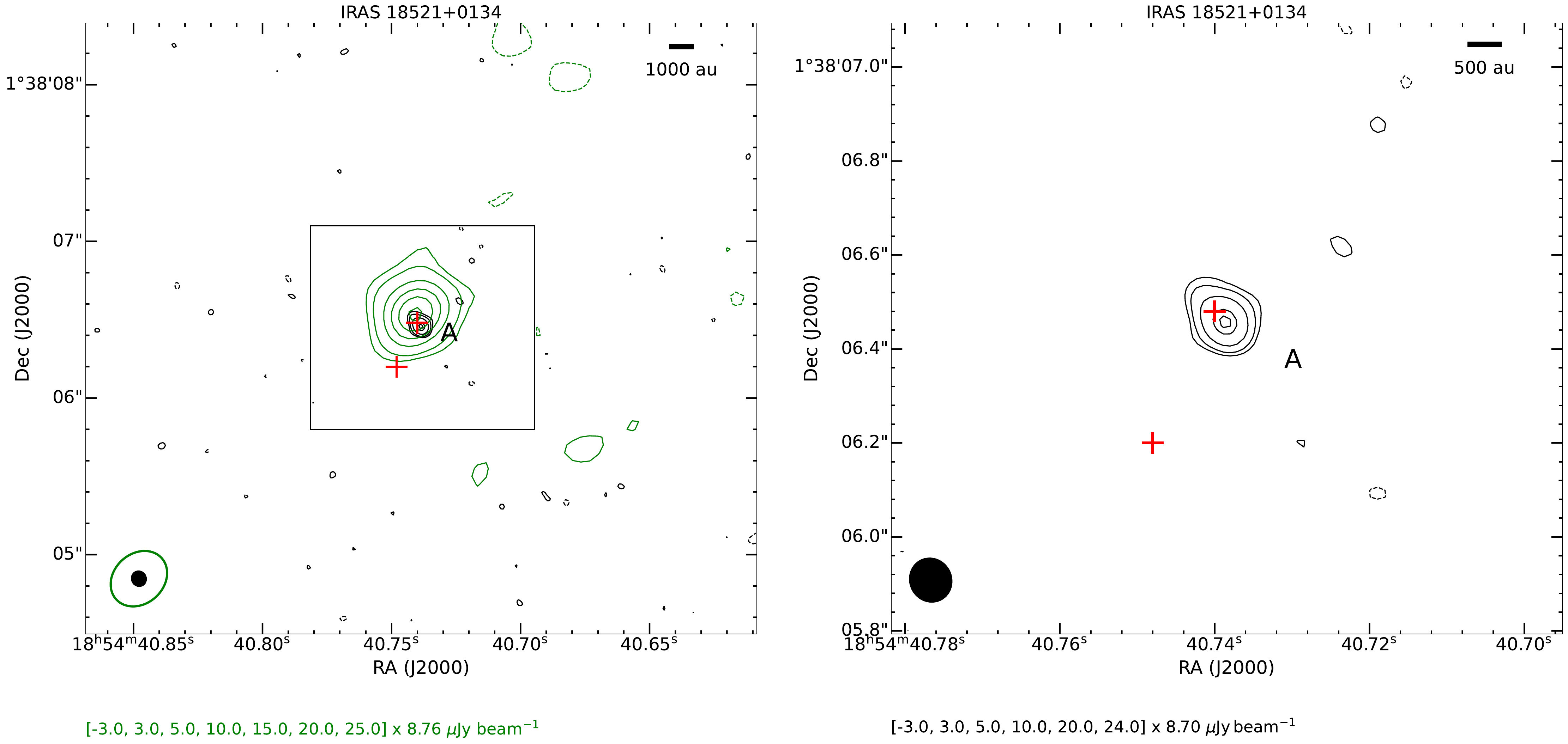}
\includegraphics[width=.9\textwidth]{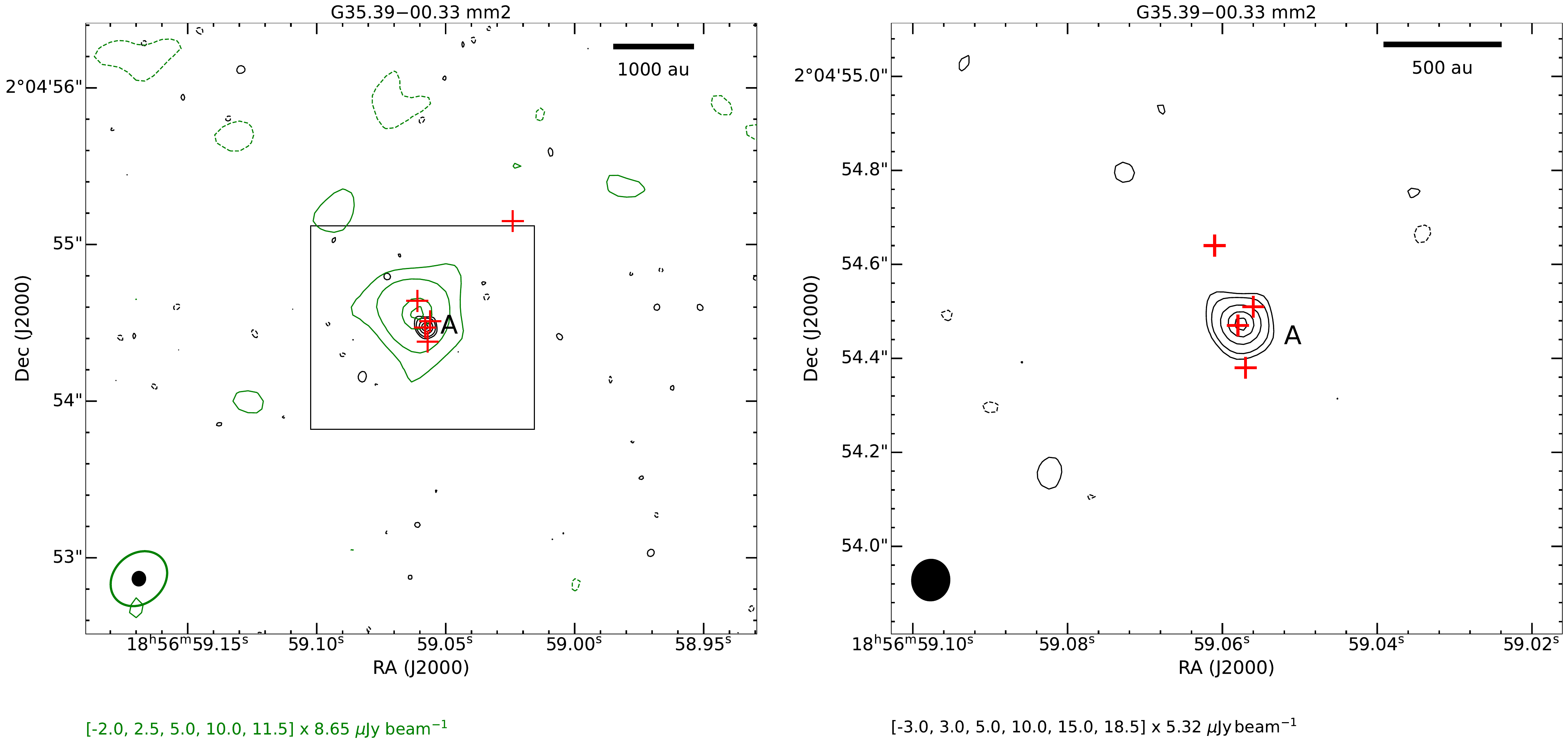}
\includegraphics[width=.9\textwidth]{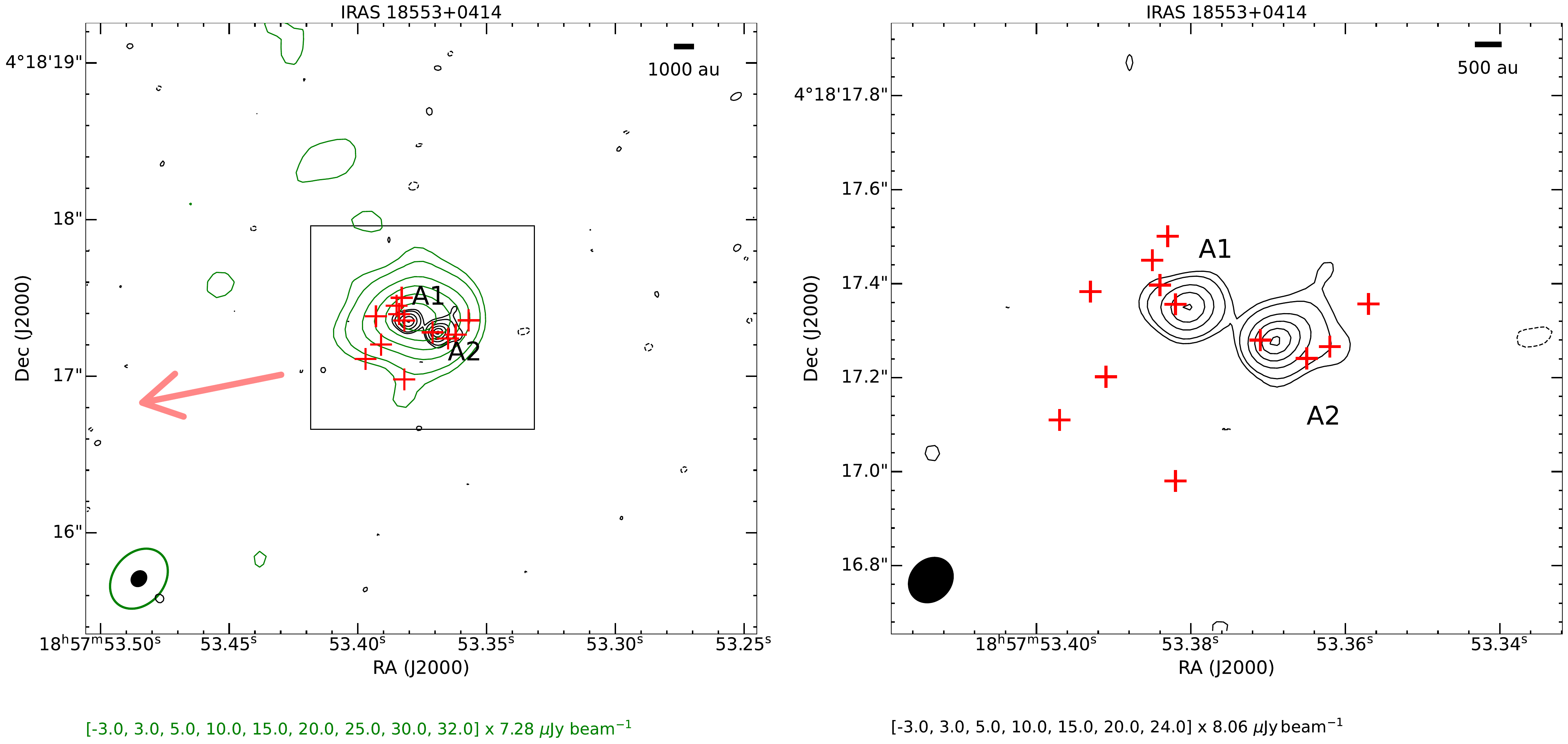}
\caption{Same as Figure \ref{fig:appendix-A1} but for IRAS 18521+0134 (top), G35.39$-$00.33 mm2 (middle), and IRAS 18553+0414 (bottom). }
\label{fig:appendix-A7}
\end{figure}

\begin{figure}[!ht]
\centering
\includegraphics[width=.9\textwidth]{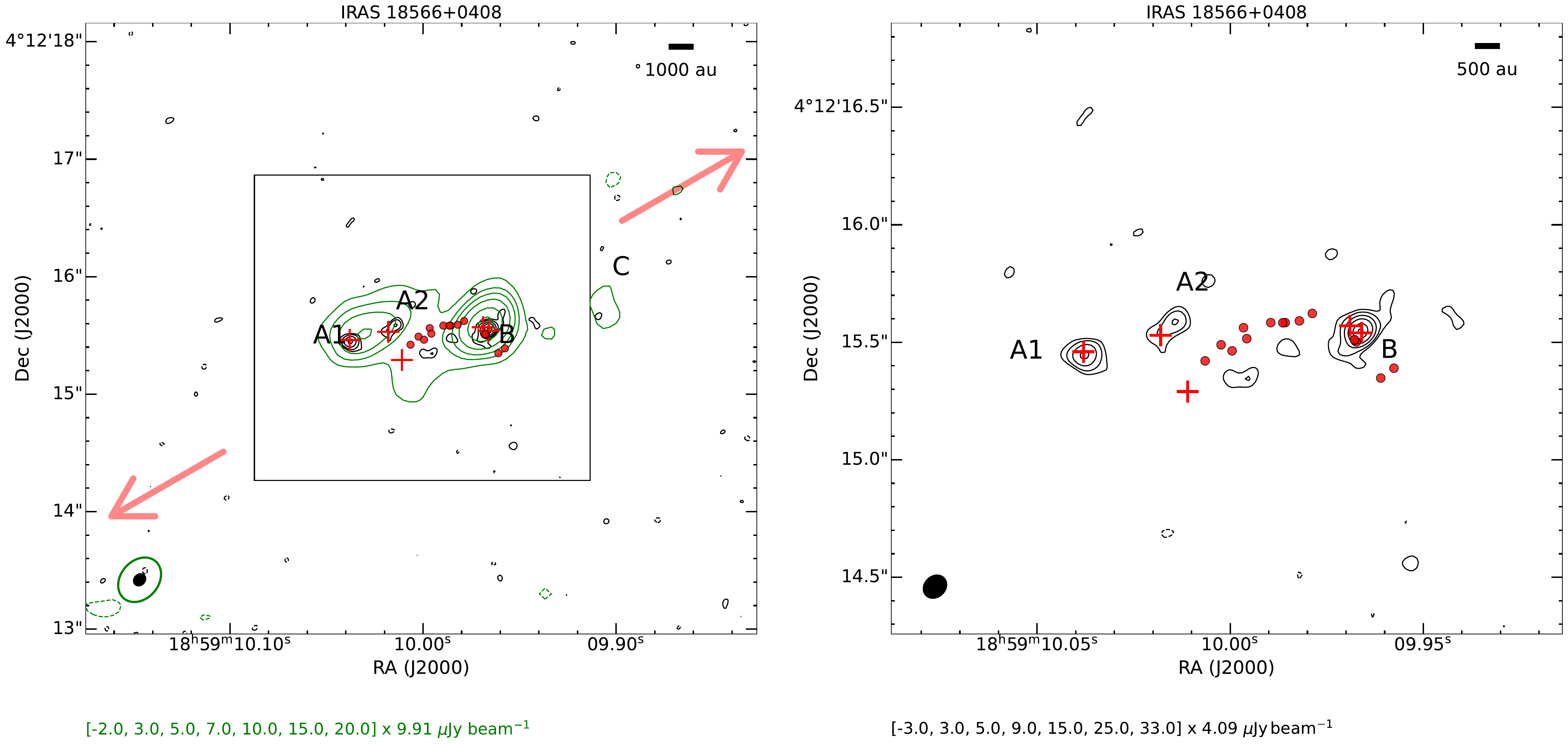}
\includegraphics[width=.9\textwidth]{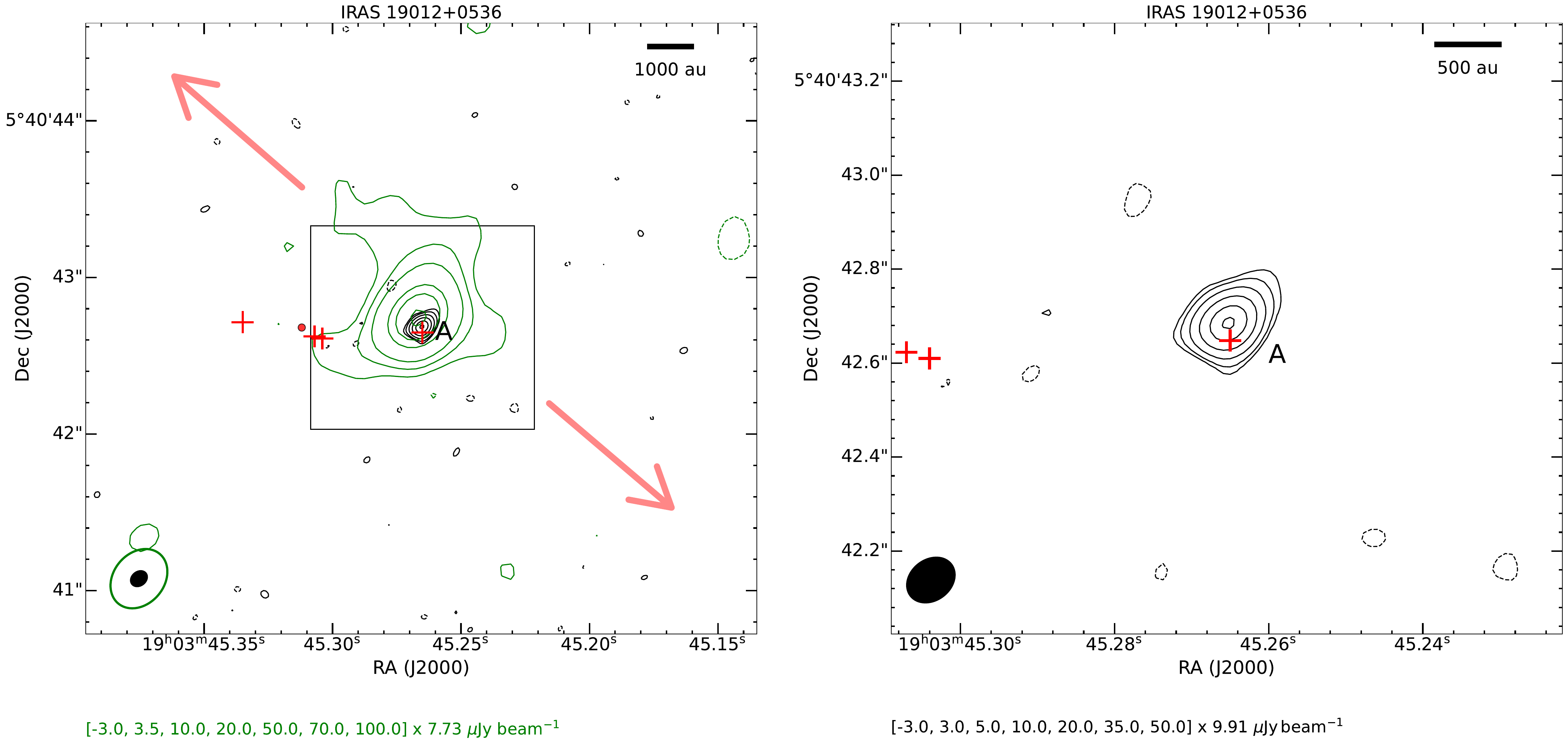}
\includegraphics[width=.9\textwidth]{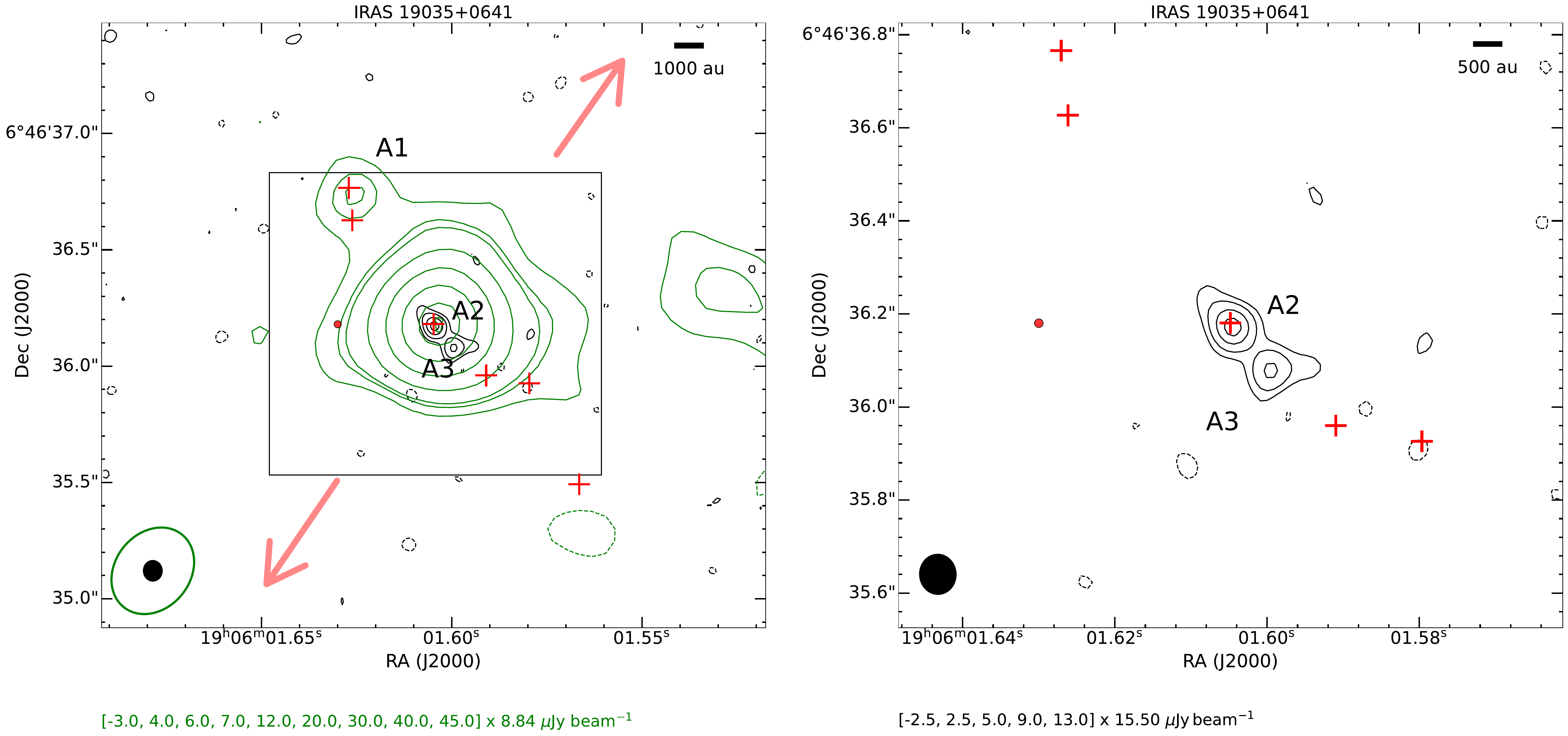}
\caption{Same as Figure \ref{fig:appendix-A1} but for IRAS 18566+0408 (top), IRAS 19012+0536 (middle), and IRAS 19035+0641 (bottom). }
\label{fig:appendix-A8}
\end{figure}

\begin{figure}[!ht]
\centering
\includegraphics[width=.9\textwidth]{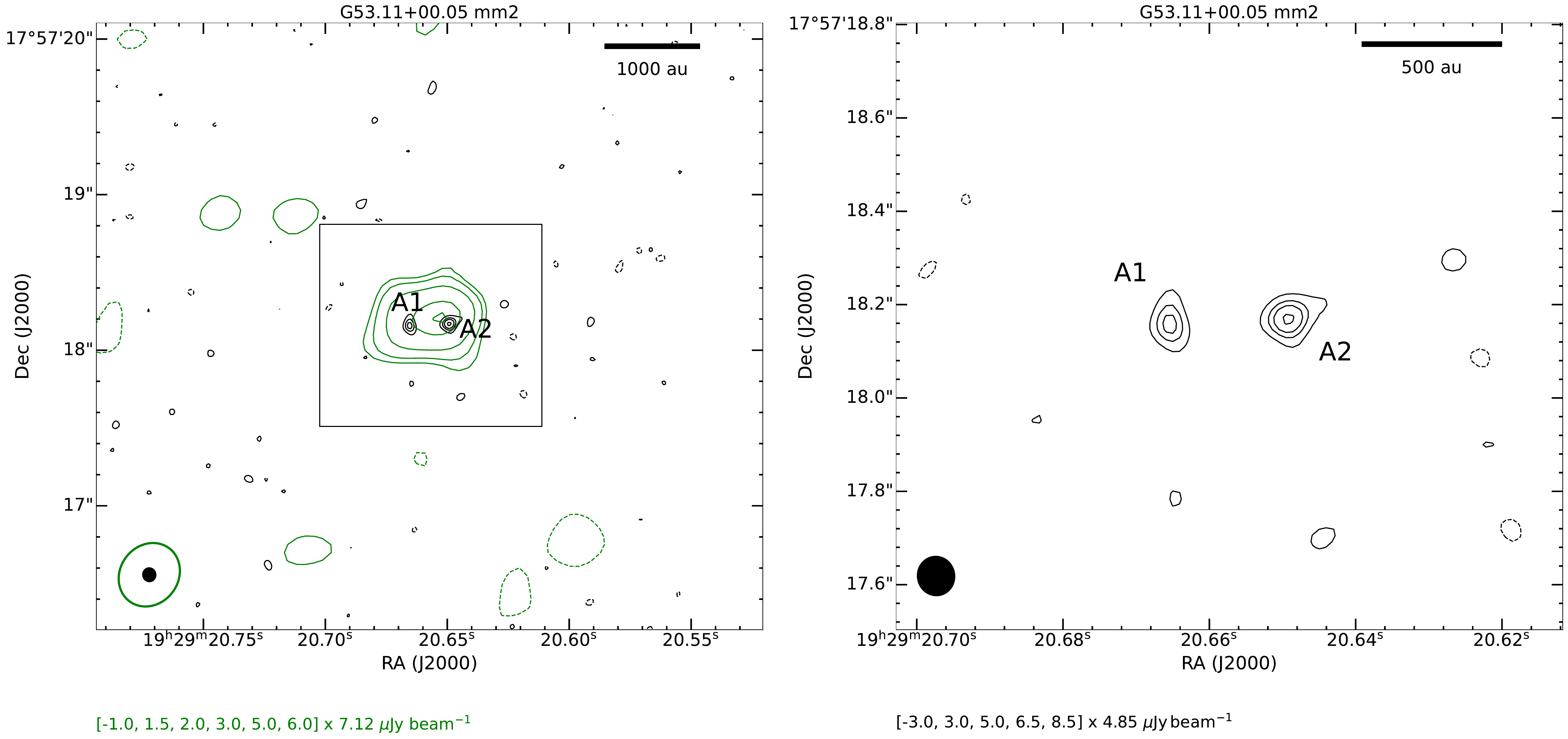}
\includegraphics[width=.9\textwidth]{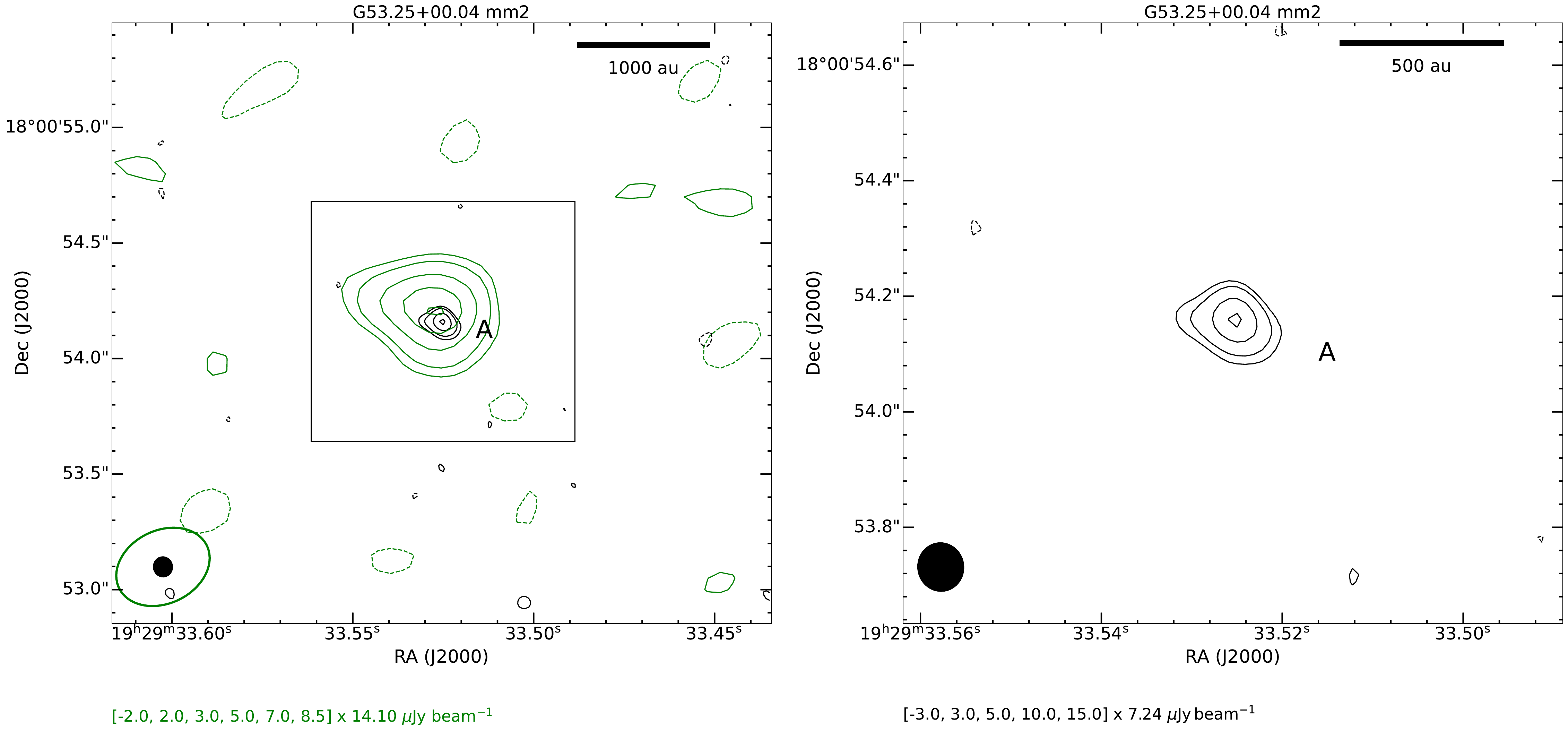}
\includegraphics[width=.9\textwidth]{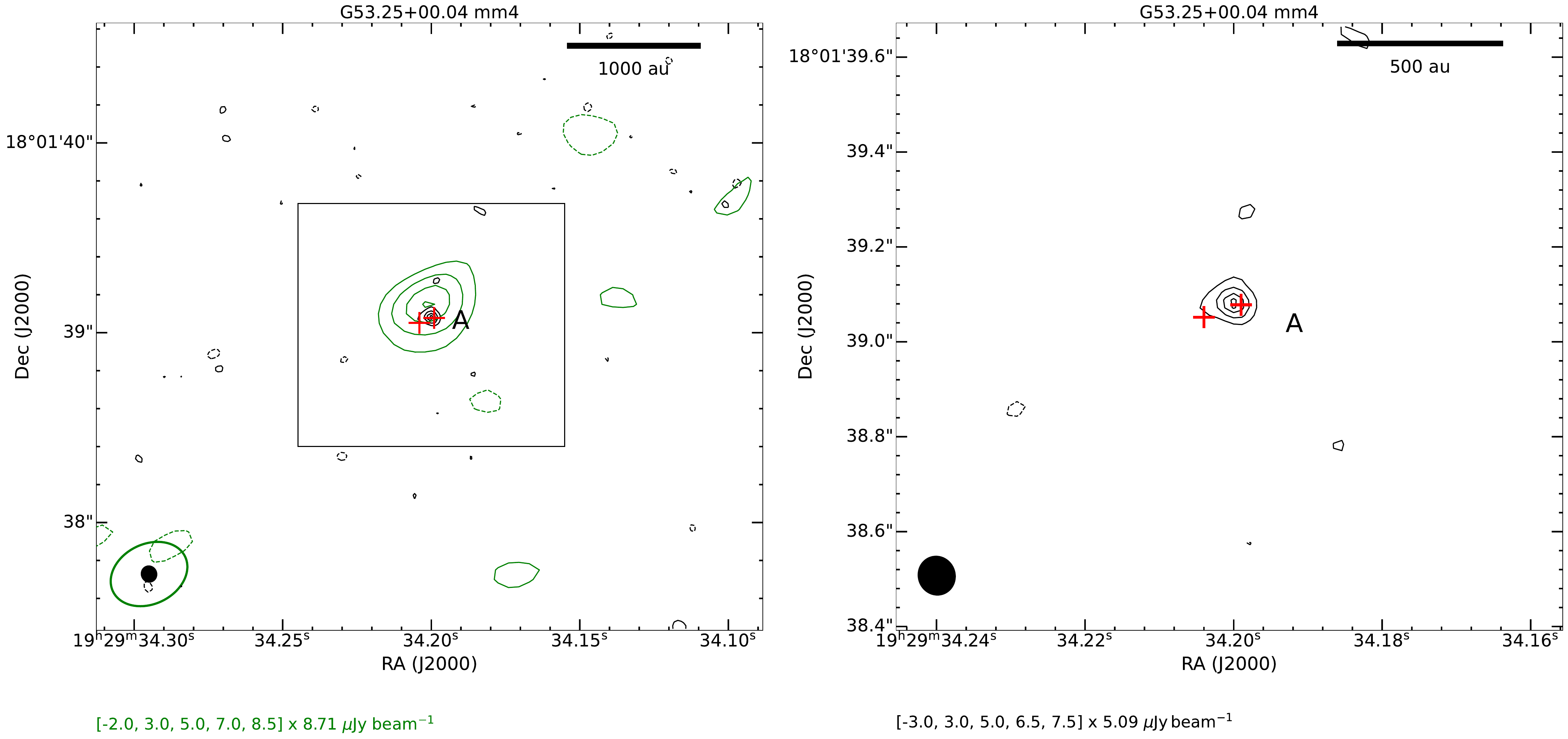}
\caption{Same as Figure \ref{fig:appendix-A1} but for G53.11+00.05 mm2 (top), G53.25+00.04 mm2 (middle), and G53.25+00.04 mm4 (bottom). }
\label{fig:appendix-A9}
\end{figure}

\begin{figure}[!ht]
\centering
\includegraphics[width=.9\textwidth]{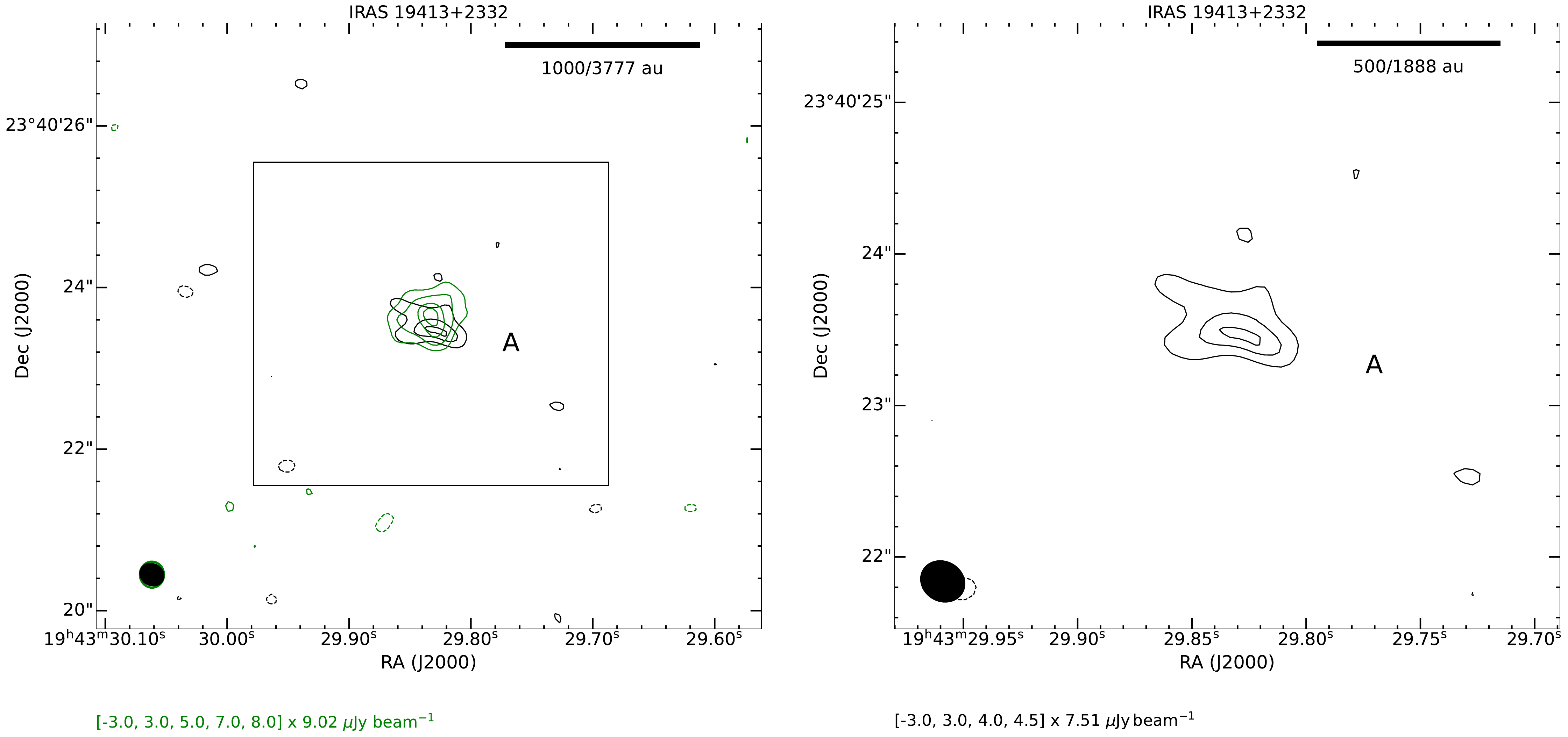}
\includegraphics[width=.9\textwidth]{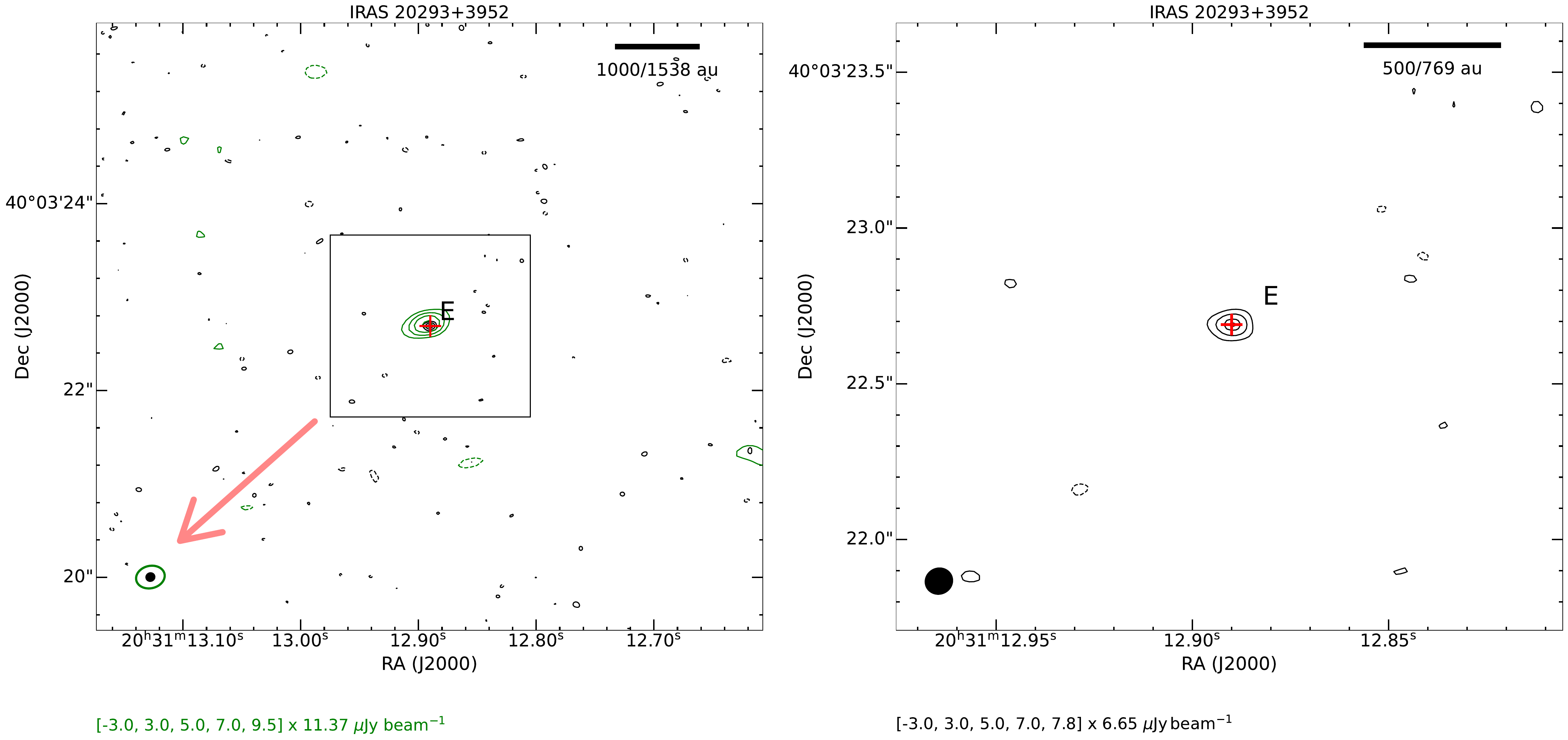}
\includegraphics[width=.9\textwidth]{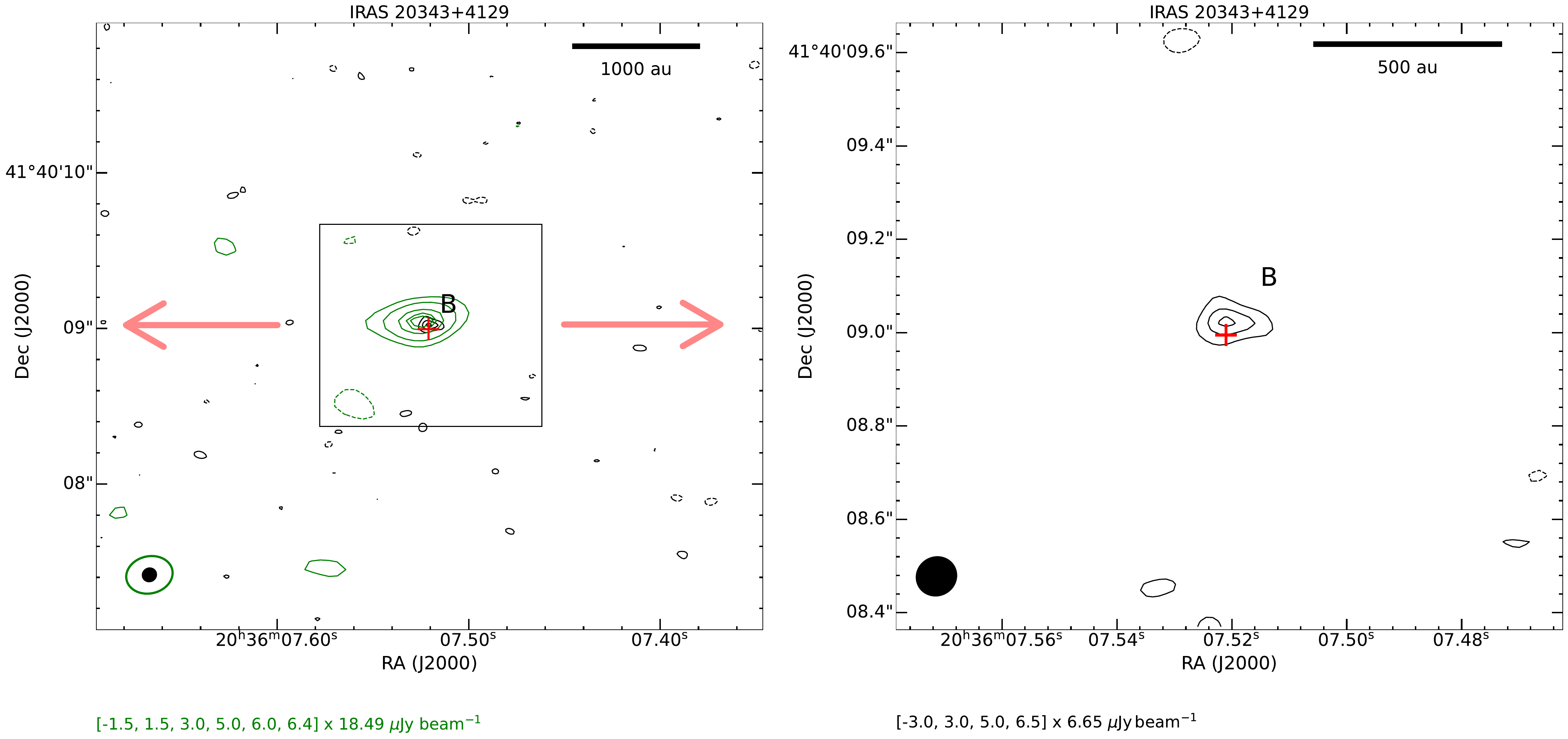}
\caption{Same as Figure \ref{fig:appendix-A1} but for IRAS 19413+2332 (top), IRAS 20293+3952 E (middle), and IRAS 20343+4129 (bottom). }
\label{fig:appendix-A10}
\end{figure}

\end{document}